\documentclass[manyauthors,nocleardouble,COMPASS]{cernphprep}
\usepackage{bm} 
\usepackage{cite}                                                         %
\newcommand{\id}{\mathrm{d}}
\usepackage{color}                                                       

\begin {document}

\begin{titlepage}
\PHnumber{2010--018}
\PHdate{\vtop to -2cm{\raggedleft 6 July 2010\\revised 16 September 2010}}

\title{
\vspace*{1cm}
Azimuthal asymmetries of charged hadrons produced by 
high-energy muons scattered off longitudinally polarised 
deuterons}  

\Collaboration{The COMPASS Collaboration}
\ShortAuthor{The COMPASS Collaboration}
\ShortTitle{Azimuthal asymmetries \ldots}

\begin{abstract}
Azimuthal asymmetries in semi-inclusive production of positive 
($h^+$) and negative hadrons ($h^-$) have been measured by 
scattering 160 GeV muons off longitudinally polarised 
deuterons at CERN. The asymmetries were decomposed in several 
terms according to their expected modulation 
in the azimuthal angle $\phi$ of the outgoing hadron. Each term receives 
contributions from one or several spin and 
transverse-momentum-dependent parton distribution and 
fragmentation functions. The amplitudes of all $\phi$-modulation 
terms of the hadron asymmetries integrated over the kinematic 
variables  are found to be consistent with zero within 
statistical errors, while the constant terms are nonzero and  
equal for $h^+$ and $h^-$ within the statistical errors. The 
dependencies of the $\phi$-modulated terms versus the Bjorken 
momentum fraction $x$, the hadron fractional momentum $z$, and 
the hadron transverse momentum $p_h^T$  were studied. The  $x$ 
dependence of the constant terms for both positive and negative 
hadrons is in agreement
with the longitudinal double-spin hadron asymmetries, measured in 
semi-inclusive deep-inelastic scattering. The $x$ dependence of the 
$\sin\phi$-modulation term is less pronounced than that in the 
corresponding HERMES data. All other dependencies of the 
$\phi$-modulation amplitudes are consistent with zero within the 
statistical errors.
\end{abstract}

\vspace*{60pt}
\begin{flushleft}
\end{flushleft}

\vfill
\Submitted{(Submitted to the European Physical Journal C)}
\end{titlepage}

{\pagestyle{empty}
%
%

\section*{The COMPASS Collaboration}
\label{app:collab}

\begin{flushleft}
M.G.~Alekseev\Irefn{turin_i},
V.Yu.~Alexakhin\Irefn{dubna},
Yu.~Alexandrov\Irefn{moscowlpi},
G.D.~Alexeev\Irefn{dubna},
A.~Amoroso\Irefn{turin_u},
A.~Austregesilo\Irefnn{cern}{munichtu},
B.~Bade{\l}ek\Irefn{warsaw},
F.~Balestra\Irefn{turin_u},
J.~Barth\Irefn{bonnpi},
G.~Baum\Irefn{bielefeld}
% ...
\end{flushleft}

%
%

\begin{Authlist}
\item \Idef{bielefeld}{Universit\"at Bielefeld, Fakult\"at f\"ur Physik, 33501 Bielefeld, Germany\Arefs{f}}
\item \Idef{bochum}{Universit\"at Bochum, Institut f\"ur Experimentalphysik, 44780 Bochum, Germany\Arefs{f}}
\item \Idef{bonniskp}{Universit\"at Bonn, Helmholtz-Institut f\"ur  Strahlen- und Kernphysik, 53115 Bonn, Germany\Arefs{f}}
\item \Idef{bonnpi}{Universit\"at Bonn, Physikalisches Institut, 53115 Bonn, Germany\Arefs{f}}
\item \Idef{brno}{Institute of Scientific Instruments, AS CR, 61264 Brno, Czech Republic\Arefs{g}}
\item \Idef{calcutta}{Matrivani Institute of Experimental Research \& Education, Calcutta-700 030, India\Arefs{h}}
\item \Idef{dubna}{Joint Institute for Nuclear Research, 141980 Dubna, Moscow region, Russia\Arefs{i}}
\item \Idef{erlangen}{Universit\"at Erlangen--N\"urnberg, Physikalisches Institut, 91054 Erlangen, Germany\Arefs{f}}
\item \Idef{freiburg}{Universit\"at Freiburg, Physikalisches Institut, 79104 Freiburg, Germany\Arefs{f}}
\item \Idef{cern}{CERN, 1211 Geneva 23, Switzerland}
\item \Idef{liberec}{Technical University in Liberec, 46117 Liberec, Czech Republic\Arefs{g}}
\item \Idef{lisbon}{LIP, 1000-149 Lisbon, Portugal\Arefs{j}}
\item \Idef{mainz}{Universit\"at Mainz, Institut f\"ur Kernphysik, 55099 Mainz, Germany\Arefs{f}}
\item \Idef{miyazaki}{University of Miyazaki, Miyazaki 889-2192, Japan\Arefs{k}}
\item \Idef{moscowlpi}{Lebedev Physical Institute, 119991 Moscow, Russia}
\item \Idef{munichlmu}{Ludwig-Maximilians-Universit\"at M\"unchen, Department f\"ur Physik, 80799 Munich, Germany\AAref{f}{l}}
\item \Idef{munichtu}{Technische Universit\"at M\"unchen, Physik Department, 85748 Garching, Germany\AAref{f}{l}}
\item \Idef{nagoya}{Nagoya University, 464 Nagoya, Japan\Arefs{k}}
\item \Idef{praguecu}{Charles University in Prague, Faculty of Mathematics and Physics, 18000 Prague, Czech Republic\Arefs{g}}
\item \Idef{praguectu}{Czech Technical University in Prague, 16636 Prague, Czech Republic\Arefs{g}}
\item \Idef{protvino}{State Research Center of the Russian Federation, Institute for High Energy Physics, 142281 Protvino, Russia}
\item \Idef{saclay}{CEA IRFU/SPhN Saclay, 91191 Gif-sur-Yvette, France}
\item \Idef{telaviv}{Tel Aviv University, School of Physics and Astronomy, 69978 Tel Aviv, Israel\Arefs{m}}
\item \Idef{triest_i}{Trieste Section of INFN, 34127 Trieste, Italy}
\item \Idef{triest}{University of Trieste, Department of Physics and Trieste Section of INFN, 34127 Trieste, Italy}
\item \Idef{triestictp}{Abdus Salam ICTP and Trieste Section of INFN, 34127 Trieste, Italy}
\item \Idef{turin_u}{University of Turin, Department of Physics and Torino Section of INFN, 10125 Turin, Italy}
\item \Idef{turin_i}{Torino Section of INFN, 10125 Turin, Italy}
\item \Idef{turin_p}{University of Eastern Piedmont, 1500 Alessandria,  and Torino Section of INFN, 10125 Turin, Italy}
\item \Idef{warsaw}{So{\l}tan Institute for Nuclear Studies and University of Warsaw, 00-681 Warsaw, Poland\Arefs{n} }
\item \Idef{warsawtu}{Warsaw University of Technology, Institute of Radioelectronics, 00-665 Warsaw, Poland\Arefs{n} }
\item \Idef{yamagata}{Yamagata University, Yamagata, 992-8510 Japan\Arefs{k} }

\end{Authlist}

%
%
\vspace*{-\baselineskip}
\begin{Authlist}
\item \Adef{a}{Also at IST, Universidade T\'ecnica de Lisboa, Lisbon, Portugal}
\item \Adef{b}{Also at Chubu University, Kasugai, Aichi, 487-8501 Japan\Arefs{k}}
\item \Adef{c}{Also at KEK, 1-1 Oho, Tsukuba, Ibaraki, 305-0801 Japan}
\item \Adef{d}{On leave of absence from JINR Dubna}
\item \Adef{e}{Also at GSI mbH, Planckstr.\ 1, D-64291 Darmstadt, Germany}
\item \Adef{f}{Supported by the German Bundesministerium f\"ur Bildung und Forschung}
\item \Adef{g}{Suppported by Czech Republic MEYS grants ME492 and LA242}
\item \Adef{h}{Supported by SAIL (CSR), Govt.\ of India}
%
%
\item \Adef{i}{Supported by CERN-RFBR grants 08-02-91009 and 08-02-91013}
\item \Adef{j}{\raggedright Supported by the Portuguese FCT - Funda\c{c}\~{a}o para a
             Ci\^{e}ncia e Tecnologia grants POCTI/FNU/49501/2002 and POCTI/FNU/50192/2003}
\item \Adef{k}{Supported by the MEXT and the JSPS under the Grants No.18002006, No.20540299 and No.18540281; Daiko Foundation and Yamada Foundation}
\item \Adef{l}{Supported by the DFG cluster of excellence `Origin and Structure of the Universe' (www.universe-cluster.de)}
\item \Adef{m}{Supported by the Israel Science Foundation, founded by the Israel Academy of Sciences and Humanities}
\item \Adef{n}{Supported by Ministry of Science and Higher Education grant 41/N-CERN/2007/0}
\item [{\makebox[2mm][l]{\textsuperscript{*}}}] Deceased
\end{Authlist}
    
\clearpage
}

\setcounter{page}{1}
\section{Introduction}

Starting from the first polarised lepton scattering experiments 
at SLAC \cite{Alguard:1976bm} and at CERN by the EMC  
\cite{Ashman:1987hv}, the longitudinal spin structure of the 
nucleon has been investigated over the past 20 years by the SMC 
\cite{Adeva:1998vv}, HERMES \cite{Airapetian:2004zf}, CLAS 
\cite{Yun:2002td} and COMPASS \cite{Ageev:2005gh,Alekseev:2007vi} 
Collaborations. The cross-section asymmetries $A_1$ and $A_1^h$ 
were measured respectively in inclusive Deep Inelastic Scattering 
(DIS) 
\begin{equation}
\label{eq1a} \vec{\ell}+\vec{N}\to\ell' + X
\end{equation}
and Semi-Inclusive Deep-Inelastic Scattering (SIDIS)
\begin{equation}
\label{eq1b} \vec{\ell}+\vec{N}\to\ell' + h + X
\end{equation}
of longitudinally polarised leptons ($\vec{\ell}$) off 
longitudinally polarised nucleons ($\vec{N}$). SIDIS, where in 
addition to the scattered lepton a hadron $h$ is detected, gives 
access to the individual quark spin distributions. From the 
measured spin-dependent asymmetries the contributions of quark 
spins to the spin of nucleons as well as the spin quark 
distribution functions for valence and sea quarks have been 
determined. 

The hadron transverse momentum leads to a dependence of the SIDIS 
cross-section on the hadron azimuthal angle $\phi$ 
(Fig.~\ref{fig1}a) and to asymmetries in the hadron
production. The asymmetries have been predicted 
\cite{Sivers:1989cc,Collins:1992kk,Kotzinian:1994dv} and 
elaborated on in a number of theoretical papers (see 
Refs.~\cite{Kotzinian:1995cz,nine,ten,Avakian:2010br,Anselmino:2000mb,Bacchetta:2006tn} 
and references therein). The azimuthal asymmetries are related to 
transverse-momentum-dependent Parton Distribution Functions (PDF) 
and polarised and nonpolarised Parton Fragmentation Functions 
(PFF). They can depend on the transverse or longitudinal 
component of the quark spin. These asymmetries were first 
observed by SMC \cite{Bravar:1999rq}, HERMES 
\cite{Airapetian:1999tv} and CLAS \cite{Avakian:2003pk}. Further 
studies were performed both with transversely polarised targets 
by HERMES (proton) \cite{Airapetian:2004tw} and COMPASS 
(deuteron, proton) \cite{Alexakhin:2005iw,Alekseev:2010rw} and 
with longitudinally polarised  targets by HERMES (proton, 
deuteron) \cite{Airapetian:2005jc,Airapetian:2002mf} and CLAS 
(proton) \cite{Avakian:2010ae}. Some of the measured azimuthal 
asymmetries (e.g.\ the so-called ``Collins asymmetry") for the 
proton are rather large, reaching up to 10\% 
\cite{Airapetian:2004tw}, others however (e.g.\ the so-called 
``pretzelosity"), do not exceed a couple of percent 
\cite{Kotzinian:2007uv} and their investigation requires very 
high statistics. The asymmetries for the deuteron are found to be 
much smaller than those for the proton indicating opposite signs 
of contributions from $u$ and $d$ quarks 
\cite{Avakian:2010br,Anselmino:2000mb}.

A search for azimuthal asymmetries for unidentified hadrons
with longitudinally polarised deuterons is presented below. These 
new data will test in a wide $(x,Q^2)$ range the existence of 
still unobserved azimuthal asymmetries connected to several 
transverse-momentum-dependent PDFs.

The Paper is organised as follows. In Section~2 a short 
theoretical overview with the basic formulae is given. The 
analysis method and the data selection are described in Sections~3
and 4, respectively. The results are presented in Section~5 and 
their stability and systematic uncertainty are discussed in 
Section~6, followed by the conclusions in Section~7. 

\section{Theoretical framework}

\begin{figure}[tb]
\centering
\begin{tabular}{cc}
\raisebox{40mm}{\rotatebox{-90}{
\includegraphics[width=.25\textwidth]{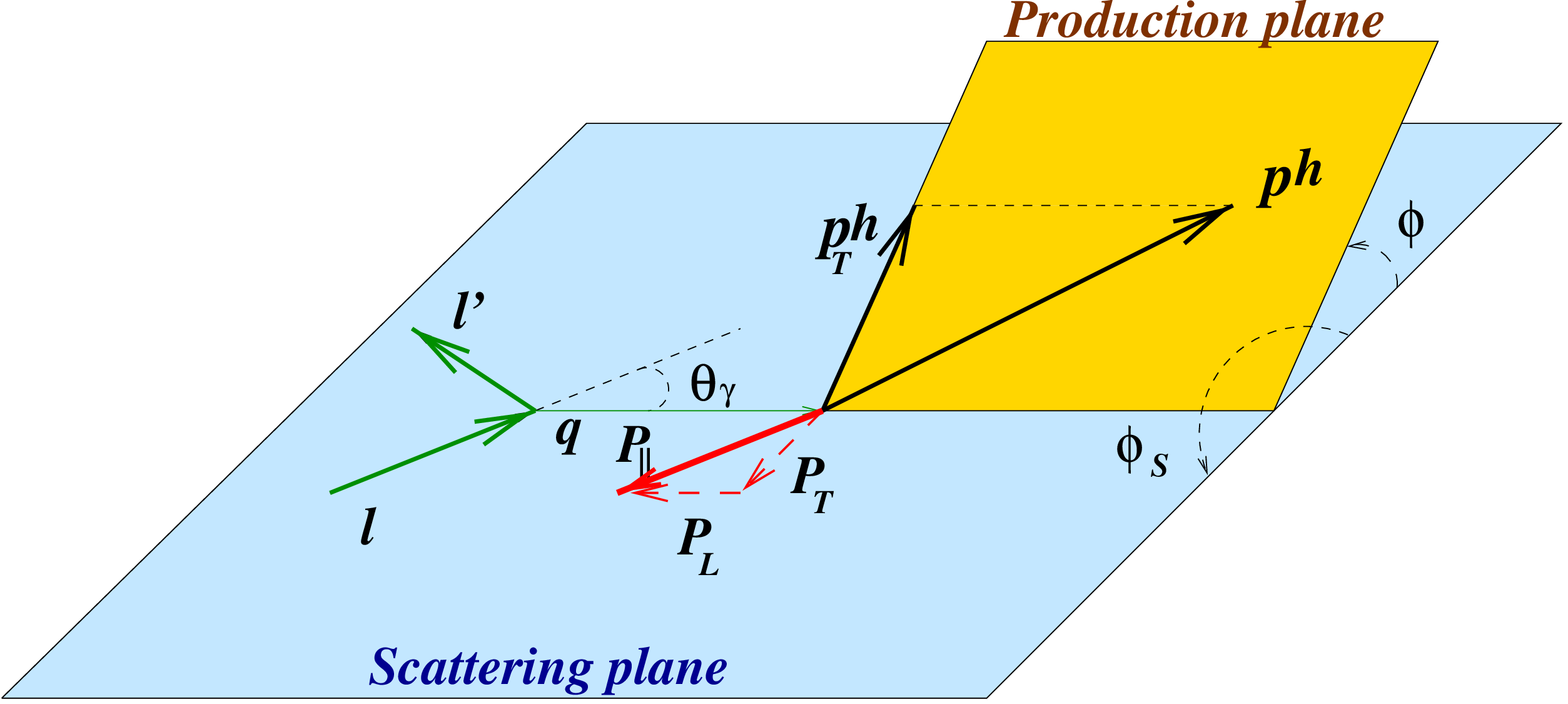}}}&
\includegraphics[width=.30\textwidth]{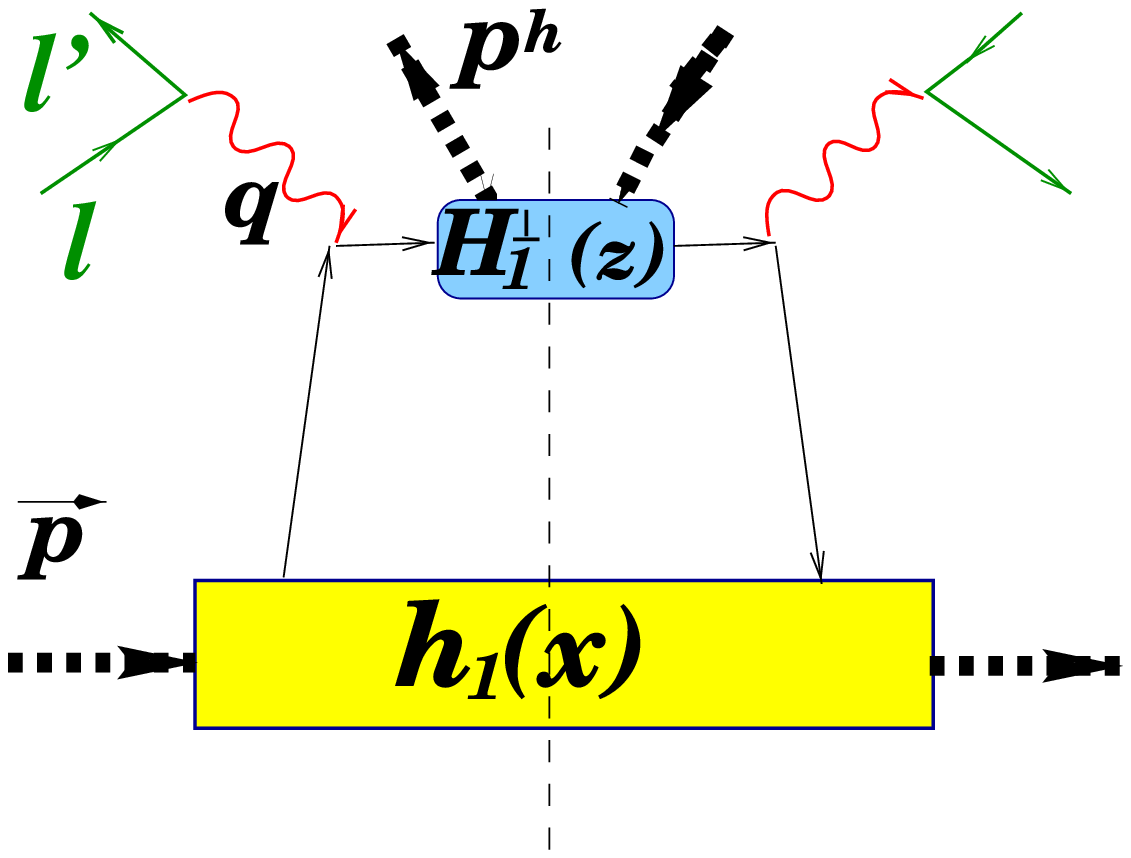}\cr
 {\bf(a)} & {\bf(b)}
\end{tabular}
\caption{\label{fig1}
{(a)} The kinematics of the SIDIS process shown for antiparallel target
polarisation $P_\|$ w.r.t.\ the beam momentum 
($\phi_S=\pi$). 
{(b)} Squared modulus of the matrix element of the SIDIS reaction 
$\ell+\vec{N}\to\ell'+h+X$ summed over the final states $X$.}
\end{figure}

The kinematics of the SIDIS reaction is shown in 
Fig.~\ref{fig1}a. The 4-momenta of the incident and scattered 
muon are denoted by $\ell$ and $\ell'$, respectively. The 
momentum transfer is given by $q=\ell-\ell'$ with $Q^2=-q^2$ and 
$\theta_\gamma$ is the angle of the momentum $\vec {q}$ of the 
virtual photon with respect to the beam. The vectors $p^h$ and 
$P_\|$ are the hadron momentum and the longitudinal target 
polarisation, respectively. Their transverse components $p_T^h$ 
and $P_T$ are defined with respect to the virtual photon 
momentum. The longitudinal component $|P_L|= 
P_\|\cos\theta_\gamma$ is approximately equal to $P_\|$ due to 
the smallness of the angle $\theta_\gamma$. The small transverse 
component is equal to 
$\left|{P_T}\right|=P_{\|}\sin(\theta_\gamma)$ where 
$\sin(\theta_\gamma)\approx 2({Mx}/{Q})\sqrt{1 - y}$, $M$ is the 
nucleon mass and $y= {q\cdot p}/{p\cdot\ell}$ is the fraction of 
the muon energy lost in the laboratory reference frame. The angle
$\phi$ is the azimuthal angle between the scattering plane and
the hadron production plane and $\phi_S$ is the angle of the 
target polarisation vector with respect to the lepton scattering 
plane. The invariant mass squared of the virtual photon--nucleon 
system $W^2$, the Bjorken variable $x$ and the hadron momentum 
fraction $z$, characterising SIDIS together with $Q^2$ and $y$, 
are defined as $W^2=(p+q)^2$, $x=Q^2/2p\cdot q$, $z=p\cdot 
p^h/p\cdot q$ where $p$ is the 4-momentum of the incident 
nucleon. 

The general expression of the total differential cross-section 
for the SIDIS reaction is a linear function of the muon beam 
polarisation $P_\mu$ and of the target polarisation components 
$P_L$ and $P_T$
\begin{equation}
\label{eq1} \id\sigma=\id\sigma_{00}+P_\mu \id\sigma_{L0}+P_L\left( 
{\id\sigma_{0L} +P_\mu \id\sigma_{LL}}\right)+\left|{P_T} 
\right|\left({\id\sigma_{0T}+P_\mu \id\sigma_{LT}}\right)
\end{equation}
where the first (second) subscript of the partial cross-sections 
refers to the beam (target) polarisation.

The asymmetry $a(\phi)$ for hadron production from a 
longitudinally polarised target is defined by
\begin{equation}
\label{eq2} 
a(\phi)=\frac{\id\sigma^{\leftarrow\Rightarrow}-\id\sigma^{\leftarrow\Leftarrow}} 
{|P_L|(\id\sigma^{\leftarrow\Rightarrow}+\id\sigma^{\leftarrow\Leftarrow})}  
=-\frac{\id\sigma_{0L}+P_\mu \id\sigma_{LL} 
-\tan\theta_\gamma\left({\id\sigma_{0T}+P_\mu 
\id\sigma_{LT}}\right)}{\id\sigma_{00}+P_\mu \id\sigma_{L0}}
\end{equation}
where $\Rightarrow$ or $\Leftarrow$ denotes the target 
polarisation along or opposite to the muon beam direction and 
$\leftarrow$ denotes the beam polarisation, which is always 
opposite to the beam direction.
The partial cross-sections $\id\sigma_{00}$ and
$\id\sigma_{L0}$ do not contribute to the numerator of the
asymmetry (Eq.~(\ref{eq2})) while $\id\sigma_{0T}$ and
$\id\sigma_{LT}$ are suppressed by the small value of
$|P_T|/|P_L|=\tan\theta_\gamma\approx 2({Mx}/{Q})\sqrt{1 - y}$.

In the parton model (one-photon approximation, handbag-type
diagram) the squared modulus of the matrix element of the SIDIS
reaction is represented by a diagram of the type shown in
Fig.~\ref{fig1}b.  The chiral-odd transversity\footnote{In this 
Paper we follow the Amsterdam 
notations\cite{nine,Bacchetta:2006tn} for PDF and PFF.} PDF 
$h_{1}(x)$ coupled to the chiral-odd Collins PFF $H_1^\bot(z)$ is
given as an example. Each of the partial cross-sections in
Eq.~(\ref{eq1}) is characterised by several terms including a
convolution of PDF and PFF multiplied by a function of the
azimuthal angle of the outgoing hadron. 
Ignoring the pure twist-3 (``tilde") fragmentation functions
and retaining only terms up to order $(M/Q)$, the contributions
to Eq.~(\ref{eq2}) for unpolarised or spin-zero hadron production
have the forms
\begin{eqnarray}
\id\sigma_{00} &\propto& xf_1(x)\otimes D_1(z)+\epsilon
xh_1^\bot(x)\otimes H_1^\bot(z)\cos(2\phi) \hfill\nonumber \\
&&+\sqrt{2\epsilon(1+\epsilon)}\frac{M}{Q}x^2 %
\left(h(x)\otimes H_1^\bot(z) %
+f^\bot(x)\otimes D_1(z)\right)\cos\phi,\nonumber \\
\id\sigma_{L0}&\propto& \sqrt{2\epsilon(1-\epsilon)}
{\frac{M}{Q}x^2\left(e(x)\otimes H_1^\bot(z)+
g^\bot(x)\otimes D_1^\bot(z)\right)\sin\phi},\nonumber\\
\id\sigma_{0L} &\propto& \epsilon xh_{1L}^\bot (x)\otimes
H_1^\bot(z)\sin(2\phi)\nonumber\\%
&&+\sqrt{2\epsilon(1+\epsilon)} \frac{M}{Q}x^2\left(h_L
(x)\otimes H_1^\bot(z)
+f_L^\bot(x)\otimes D_1(z)\right)\sin\phi,\label{eq3}\\
\id\sigma_{LL} &\propto& \sqrt{1-\epsilon^2}xg_{1L}(x)\otimes
D_1(z)\nonumber\\%
&&+\sqrt{2\epsilon(1-\epsilon)}\frac{M}{Q}x^2
\left(g_L^\bot(x)\otimes D_1(z)
+e_L(x)\otimes H_1^\bot(z)\right)\cos\phi,\nonumber\\
\id\sigma_{0T} &\propto& \epsilon\big\{xh_1(x)\otimes
H_1^\bot(z)\sin(\phi+\phi_S)%
+xh_{1T}^\bot(x)\otimes H_1^\bot(z)\sin(3\phi-\phi_S)\big\}
\nonumber\\%
&&+xf_{1T}^\bot (x)\otimes D_1(z)\sin(\phi-\phi_S),\nonumber\\
\id\sigma_{LT}&\propto&\sqrt{1-\epsilon^2}{xg_{1T}(x)\otimes
D_1(z)\cos(\phi-\phi_S)}\,,\nonumber
\end{eqnarray}
where the target spin angle is $\phi _{S}=0$ or $\pi$. The ratio
of the virtual-photon flux with longitudinal to that with
transverse polarisation is given by $\epsilon\approx
{2(1-y)}/({2-2y+y^2})$. The symbol $\otimes$ represents
a convolution of a PDF and a PFF weighted by a function of
transverse momenta as defined in Eqs.~(4.1--4.19) of
Ref.~\cite{Bacchetta:2006tn} where also the exact expressions of
the cross-sections are given.


Some of the asymmetry modulations arising from Eqs.~(\ref{eq2}) 
and (\ref{eq3}) have already been observed in experiments 
with either transversely or longitudinally polarised targets.
These asymmetries involve the $\sin\phi$ modulation in 
$\id\sigma_{0L}$ due to the twist-3 PDFs $h_L$ and $f_L^\perp$
and that in $\id\sigma_{0T}$ due to transversity $h_1$ and the
Sivers PDF $f_{1T}^\bot$. The $\sin(2\phi)$ modulation in 
$\id\sigma_{0L}$ arising from the ``worm-gear'' PDF $h_{1L}^\bot$ 
has been seen also. Other asymmetries have not yet been observed
experimentally. These include the $\sin(3\phi)$ modulation in 
$\id\sigma_{0T}$ related to the ``pretzelosity'' PDF and the 
$\cos\phi$ modulations both in $\id\sigma_{LT}$ due the 
``worm-gear'' PDF $g_{1T}$ and in $\id\sigma_{LL}$ due to the 
twist-3 $g_L^\bot$ and $e_L$. 
 
The aim of this  study is to evaluate the azimuthal asymmetries 
in hadron production from the longitudinally polarised 
target as a manifestation of the quark-spin and 
transverse-momentum dependent PDFs and PFF mentioned above and to 
investigate the $x$, $z$ and $p_T^h$ dependence of the 
corresponding modulation amplitudes.

\section{Analysis}

The experiment was performed in the muon beam M2 at CERN with
positive 160~GeV muons. The beam is naturally polarised opposite 
to the muon momentum with an average polarisation $P_\mu=-80\%$. 

Briefly, the COMPASS setup \cite{Abbon:2007pq} is a two-stage 
forward spectrometer with the world's largest polarised target, 
tracking detectors and particle identification detectors behind 
each of the two spectrometer magnets. Various tracking detectors 
(MICROMEGAS, GEM, Straw, MWPC, DC) provide a precise 
determination of the particle coordinates, while electromagnetic 
and hadron calorimeters, muon detectors, and a RICH provide 
identification of secondary electrons, muons and hadrons. The 
fast trigger and data acquisition systems provide for high statistics 
measurements.

The method of analysis takes advantage of the polarised target 
arrangement optimised for asymmetry measurements. 
In 2002--2004 the target consisted of an upstream 
cell (``U") and a downstream cell (``D") placed along the axis of 
a 2.5~T solenoidal magnet centred along the beam direction. The 
target material $^{6}$LiD was kept at a low temperature in a 
$^3$He--$^4$He dilution refrigerator. The material in the cells 
was polarised longitudinally with opposite orientations by 
Dynamic Nuclear Polarisation. The beam traverses both cells and 
data are taken simultaneously for both polarisations. To minimise 
remaining systematic effects caused by possible time-dependent variations 
of the acceptance, the polarisation of the cells is reversed 
three times per day by adiabatically inverting the solenoid 
field. To avoid possible systematic acceptance effects resulting 
from the different solenoid field orientations, after a few weeks 
the same polarisation configuration is realised with inverted 
magnetic field.

For the azimuthal-asymmetry studies double ratios $R_{f}$ of
event numbers are used
\begin{equation}
\label{eq4} 
R_f(\phi)=\left[{N_{+,f}^U(\phi)/N_{-,f}^D(\phi)}\right]\times 
\left[{N_{+,f}^D(\phi)/N_{-,f}^U(\phi)}\right]
\end{equation}
where $N_{p,f}^t(\phi)$ is the number of events in a given $\phi$ bin 
originating from the target cell $t$ ($t=U$, $D$) with the 
polarisation orientation $p$ ($p=+$, $-$) and the 
solenoid field orientation $f$ ($f=+$, $-$) w.r.t.\ to the beam 
direction. 

Using Eqs.~(\ref{eq1}, \ref{eq3}) the number of events can be
expressed as
\begin{eqnarray}
\label{eq5} 
N_{p,f}^t =C_f^t(\phi)L_{p,f}^t
\Big[(B_0 + B_1\cos\phi + B_2\cos2\phi+B_3\sin\phi +  \dots ) \nonumber\\
\pm P_{p,f}^t(A_0  +  A_1\sin\phi  +  A_2\sin(2\phi) 
  + \dots )\Big]
\end{eqnarray}
where $C_f^t(\phi)$ is the acceptance factor, $L_{p,f}^t$ is the 
luminosity, and $P_{p,f}^t$ is the absolute value of the averaged 
product of the measured positive or negative target polarisation 
and the dilution factor\footnote{The dilution factor is given by 
the ratio of the absorption cross-sections on the deuteron to 
that of all nuclei entering the target cells. It includes a 
correction for the relative polarisation of deuterons bound in 
$^6$Li with respect to free deuterons. It also includes the 
dilution due to radiative events on the deuteron, which is taken 
into account by the ratio of the one-photon exchange 
cross-section to the total cross-section.} calculated for the 
cell $t$ \cite{Ageev:2005gh}. The coefficients $B_{0}$, 
$B_{1}$,~\ldots\ and $A_{0}$, $A_{1}$,~\ldots\ characterise  the 
target-spin-independent and the target-spin-dependent parts of 
partial cross-sections contributing to the denominator and 
numerator in Eq.~(\ref{eq2}), respectively.  

Substituting Eq.~(\ref{eq5}) into Eq.~(\ref{eq4}), one can see 
that the multiplicative acceptance factors cancel out as well as
the luminosity factors if the beam muons cross both cells. 
The ratios $R_{f}(\phi)$ thus depend only on the asymmetry 
$a(\phi)$ (Eq.~(\ref{eq2})), which can be expressed (to first 
order) by
\begin{equation}
\label{eq6} 
a_f(\phi)= \left[{R_f(\phi)-1}\right]/ 
(P_{+,f}^U+P_{+,f}^D+P_{-,f}^U+P_{-,f}^D )\,.
\end{equation}
Since the asymmetry should not depend on the orientation of the 
solenoid field, one can expect $a_{+}=a_{-}$. From the data a 
small nonzero difference between $a_{+}$ and $a_{-}$ was found. 
This difference may be due to nonfactorisable 
solenoid-field-dependent contributions in Eq.~(\ref{eq5}). 
However, these contributions have different signs, as it was 
checked by Monte Carlo simulations, and cancel out in the 
weighted sum $a(\phi)=a_{+}(\phi)\oplus a_{-}(\phi)$. Therefore, 
this sum -- calculated separately for each year of data taking 
and averaged at the end -- is used for the final results.

\section{Data selection}

The data selection starts from the full data set of SIDIS events 
taken in 2002--2004
with $Q^{2}>1$~(GeV$/c)^{2}$ and $y>0.1$. For each event a 
reconstructed vertex with incoming and outgoing muons and one or 
more additional outgoing tracks is required. Applying cuts on the 
quality of reconstructed tracks and vertices, the target volume, 
the momentum of the incoming muon ($140<|{\bm l}|<180$~GeV/c), 
the fraction of the muon energy loss ($y<0.9$) and the invariant 
mass of the virtual photon--nucleon system ($5<W<18~{\rm 
GeV}/c^2$), about $96\times10^6$ of the SIDIS events remain for 
further analysis. 

The tracks originating from SIDIS events have been identified 
as hadrons using the information from the hadron calorimeters 
HCAL1 and HCAL2 \cite{Abbon:2007pq}. A track is considered a 
hadron track if it hits one of the calorimeters, the calorimeter has an 
associated cluster with energy greater then 5~GeV in HCAL1 or 
greater then 7~GeV in HCAL2, the coordinates of the cluster are 
compatible with those of the track, and the energy of 
the cluster is compatible with the momentum of the track. The 
total number of the hadrons is about $53\times10^6$.

\begin{figure}[tb]
\centering
\includegraphics[width=0.45\textwidth]
{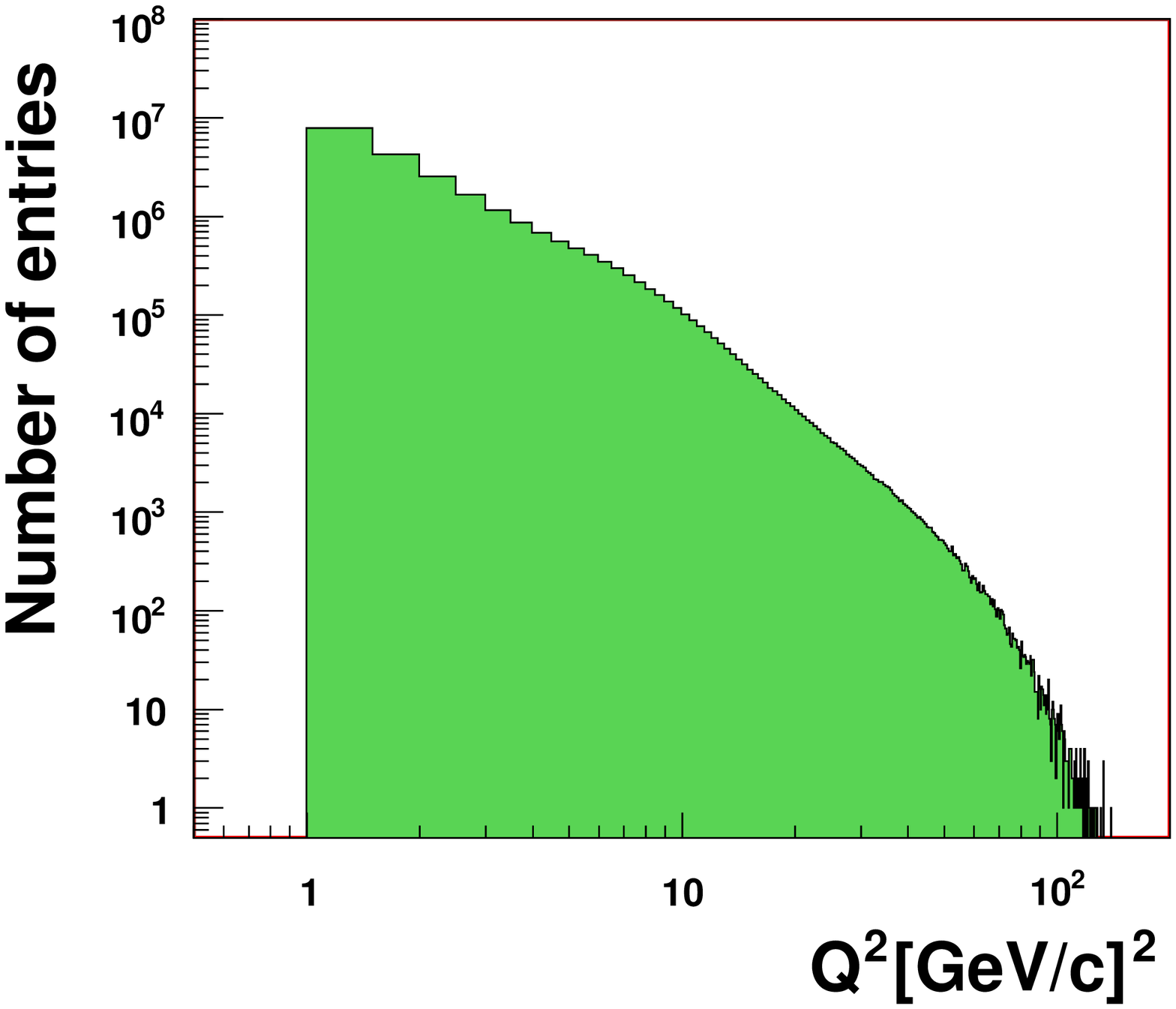}
\includegraphics[width=0.45\textwidth]
{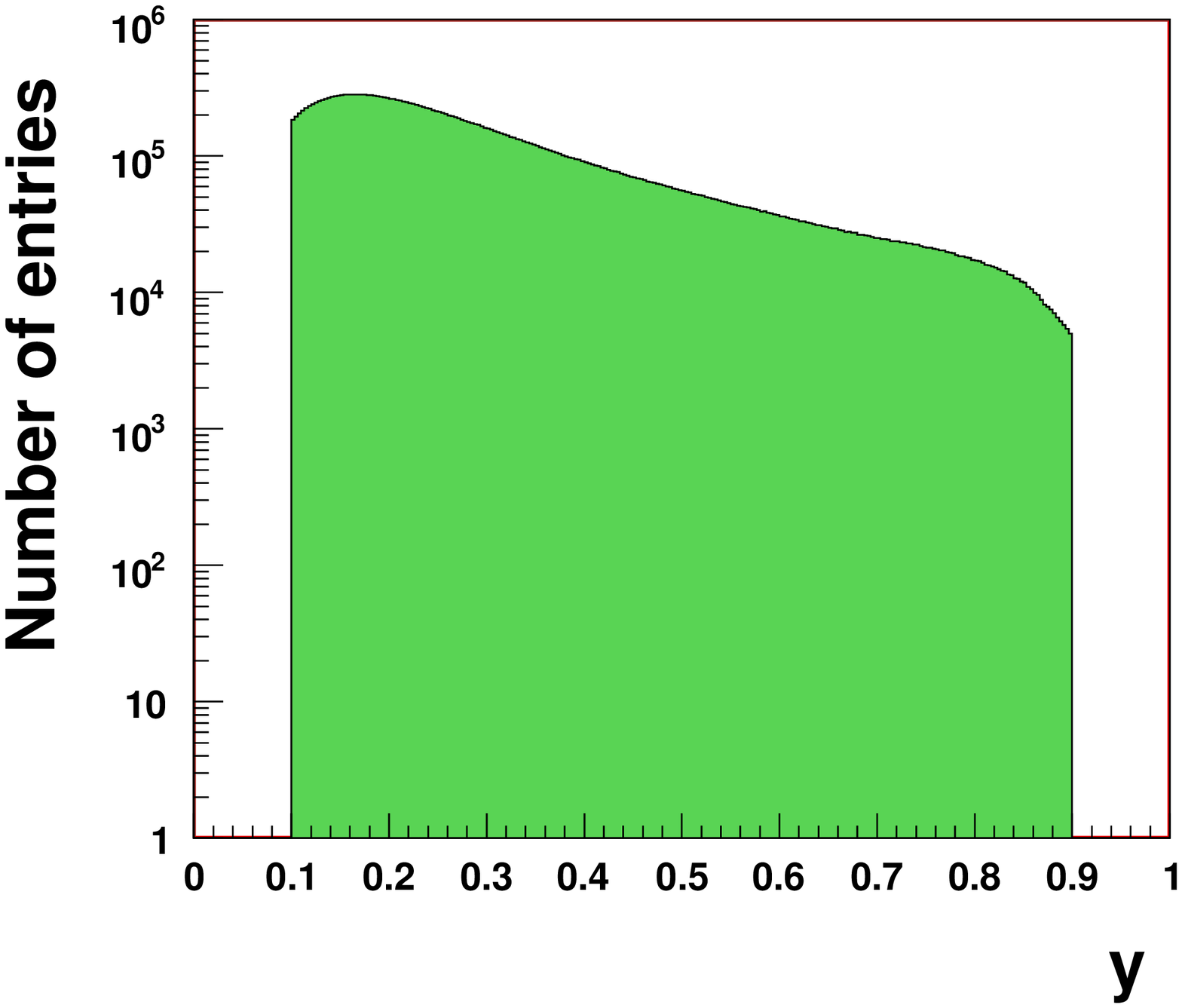}%
\vskip-3mm
\caption
{\label{fig2}Distribution of the SIDIS  events vs.\ $Q^2$ 
(left) and vs.\ $y$ (right). }
\end{figure}

\begin{figure}[tb]
\centering
\includegraphics[width=0.45\textwidth]
{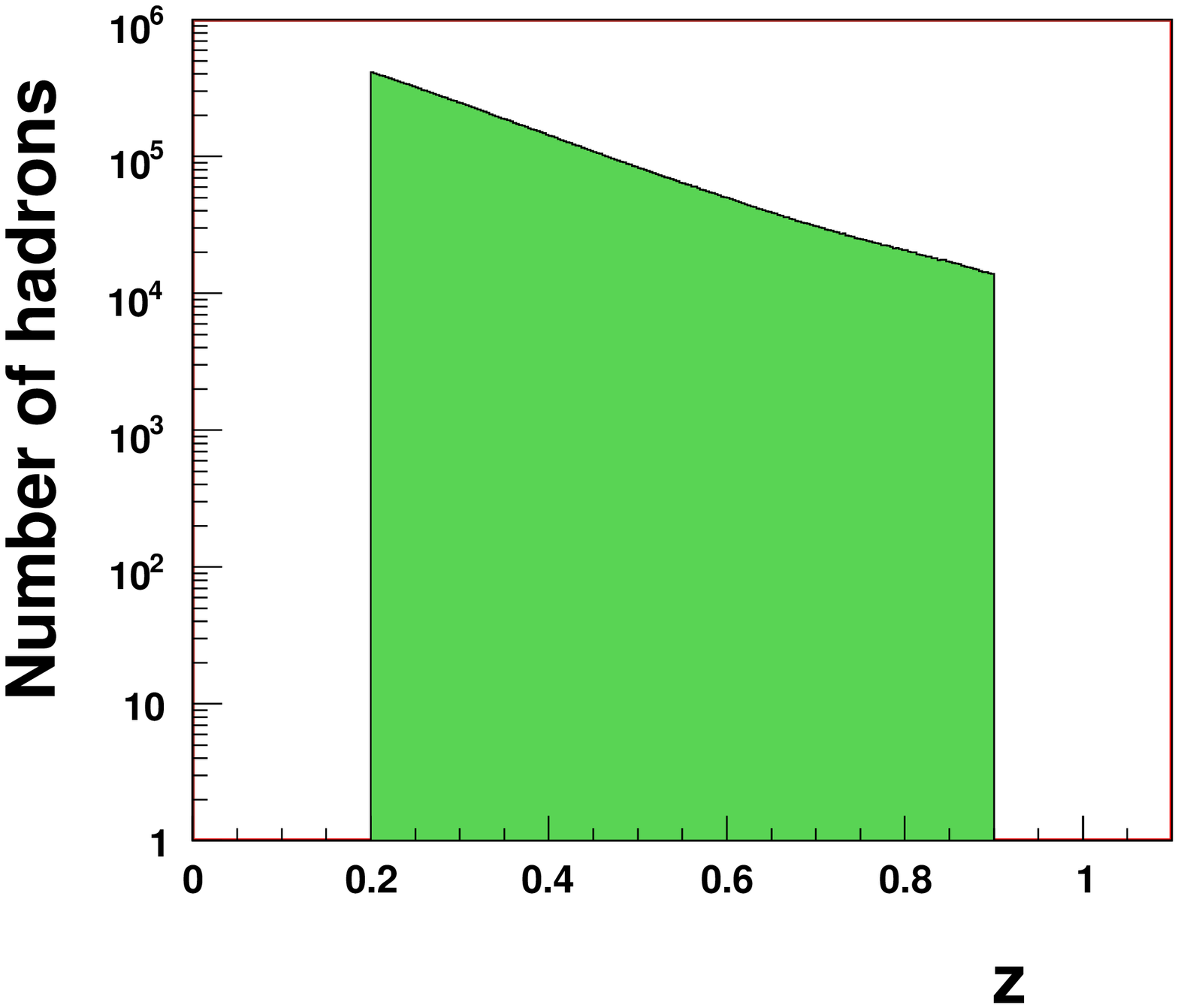}
\includegraphics[width=.45\textwidth]
{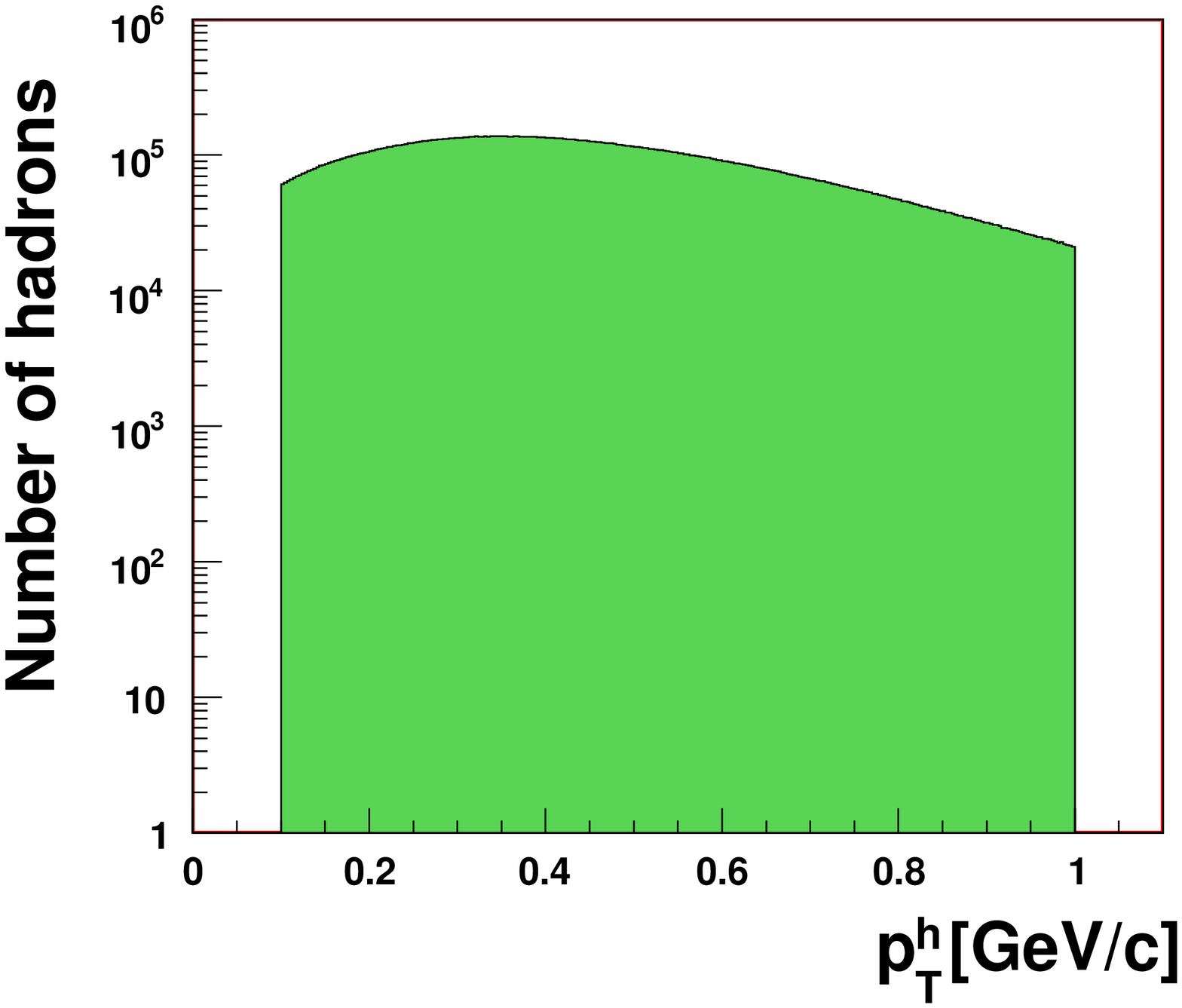}%
\vskip-3mm
\caption
{\label{fig3}Distribution of hadrons from the restricted 
kinematic region vs.\ $z$ (left) and vs.\ $p_T^h$ (right).}
\end{figure}

The asymmetries are evaluated in the restricted kinematic region 
of $x=0.004$--0.7, $z=0.2$--0.9, $p_T^h=0.1$--1~GeV/$c$. The 
lower $x$-cut value corresponds to a cut on $Q^2>1$~(GeV/$c)^2$, 
while the highest $x$ value is determined by the acceptance. The 
restriction of the energy fraction of hadrons to $z>0.2$ 
corresponds to the c.m.\ Feynman variable $x_F\approx 
z-({p^h_T}^2+m_h^2)/(zW^2)>0$ and assures that the hadron comes 
from the current fragmentation region. This cut removes almost 
half of the hadron tracks. The high-$z$ cut is applied to limit 
the influence of exclusive channels. The low-$p_T^h$ cut 
guarantees a good determination of the hadron angles while the 
high-$p_T^h$ cut corresponds kinematically to the $x_F$ and $W$ 
cuts. Distributions of the SIDIS events, from which the hadrons 
are selected for asymmetry evaluations, are shown in Fig.~\ref{fig2}
vs.\ $Q^2$ and $y$. The average value of $Q^2$ is 
3.26~(GeV/$c)^2$. The distributions of the hadrons vs.\ $z$ and 
$p_T^h$ are shown in Fig.~\ref{fig3}.

\section{Results}

The weighted sums of the azimuthal asymmetry 
$a(\phi)=a_+(\phi)\oplus a_-(\phi)$ for negative and positive 
hadrons have been fitted by the function

\begin{equation}
\label{eq7} a(\phi)=a^{\rm 
const}+a^{\sin\phi}\sin\phi+a^{\sin2\phi}\sin(2\phi)+ 
a^{\sin3\phi}\sin(3\phi)+a^{\cos\phi}\cos\phi.
\end{equation}

The fit parameters $a^{f(\phi)}$ are connected to particular PDFs 
and PFFs in Eq.~(\ref{eq3}) and can depend on $x$, $z$ and  
$p^h_T$. First, the asymmetries $a(\phi)$ integrated over these
kinematic variables have been calculated and fitted in 10 $\phi$ 
bins. In a second step the dependence of $a^{f(\phi)}$ on each of
the variables was studied while integrating over the other two 
variables. For this purpose fits of $a(\phi)$ have been performed 
in 10 bins of $\phi$ for each of the 6 bins in $x$, $z$, and $p^h_T$.
In Eq.~(\ref{eq7}) we have disregarded the contribution of
the $\cos2\phi$ term which could appear from $\id\sigma_{00}$ in 
the denominator of Eq.~(\ref{eq2}). This amplitude 
is expected\cite{Arneodo:1986cf} to not exceed 10\% and would enter 
into Eq.~(\ref{eq7}) with the factor $a^{\rm const}$ which is of
order of $10^{-3}$ (see Table~\ref{tab:1}). This is beyond our accuracy.
The same comments are valid for the contribution of the $\cos\phi$ and $\sin\phi$ amplitudes 
coming from the denominator of Eq.~(\ref{eq2}). However, their 
contributions are automatically taken into account by the corresponding fit
parameters. 

For the integrated sample the parameters characterising the 
$\phi$-modulation amplitudes are compatible with zero within 
1--1.5 standard deviations. The $\phi$-independent parameters 
$a^{\rm const}$ differ from zero and are almost equal for $h^{-}$ 
and $h^{+}$ within the statistical errors. The fit parameters are 
given in Table~\ref{tab:1}. The correlation coefficients of the 
fit parameters do not exceed 15\%. Also given are the results for 
fits considering only a constant term, $a(\phi)=a^{\rm const}$.
\begin{table}[t]
\begin{center}
\caption{Best values of the $a(\phi)$ fit parameters for positive 
         and negative hadrons from the five- and one-parameter fits.}
\label{tab:1}

\begin{tabular}{crrrr}
\hline 
\hline 
Fit parameters&\multicolumn{1}{c}{$h^-$}  &\multicolumn{1}{c}{$h^+$}&\multicolumn{1}{c}{$h^-$}&\multicolumn{1}{c}{$h^+$}\\
$\times 10^{4}$&&&&\\\hline 
$a^{\rm const}$&$23\pm17$&$~40\pm15$&$23\pm16$&$35\pm15$\\
$a^{\sin\phi} $&$15\pm23$&$-30\pm21$&\multicolumn{1}{c}{--}&\multicolumn{1}{c}{--}\\
$a^{\sin2\phi}$&$30\pm23$&$-24\pm21$&\multicolumn{1}{c}{--}&\multicolumn{1}{c}{--}\\
$a^{\sin3\phi}$&$40\pm24$&$-10\pm21$&\multicolumn{1}{c}{--}&\multicolumn{1}{c}{--}\\
$a^{\cos\phi} $&$-4\pm24$&$~38\pm22$&\multicolumn{1}{c}{--}&\multicolumn{1}{c}{--}\\
$\chi^2/$n.d.f.&   6.1/5&    1.0/5&    10.4/9&    7.0/9\\
\hline
\hline
\end{tabular}
\end{center}
\end{table}
The $\phi$-independent parts of the asymmetries appear due to the 
first term of $\id\sigma _{LL }$, which is proportional to the 
helicity PDF $g_{1L}\equiv g_1$ convoluted with the PFF of 
unpolarised quarks in an unpolarised hadron (see 
Eq.~(\ref{eq3})). For the isoscalar deuteron this contribution is 
expected to be only weakly dependent on the hadron charge, as is 
confirmed by the results. Moreover, these constants are related 
to the hadron SIDIS asymmetry $A_{1,d}^h(x)$ for deuterons (see 
below). 
  
\begin{figure}[p]
\centering 
\hspace{-5mm}
\includegraphics[width=0.33\textwidth]
{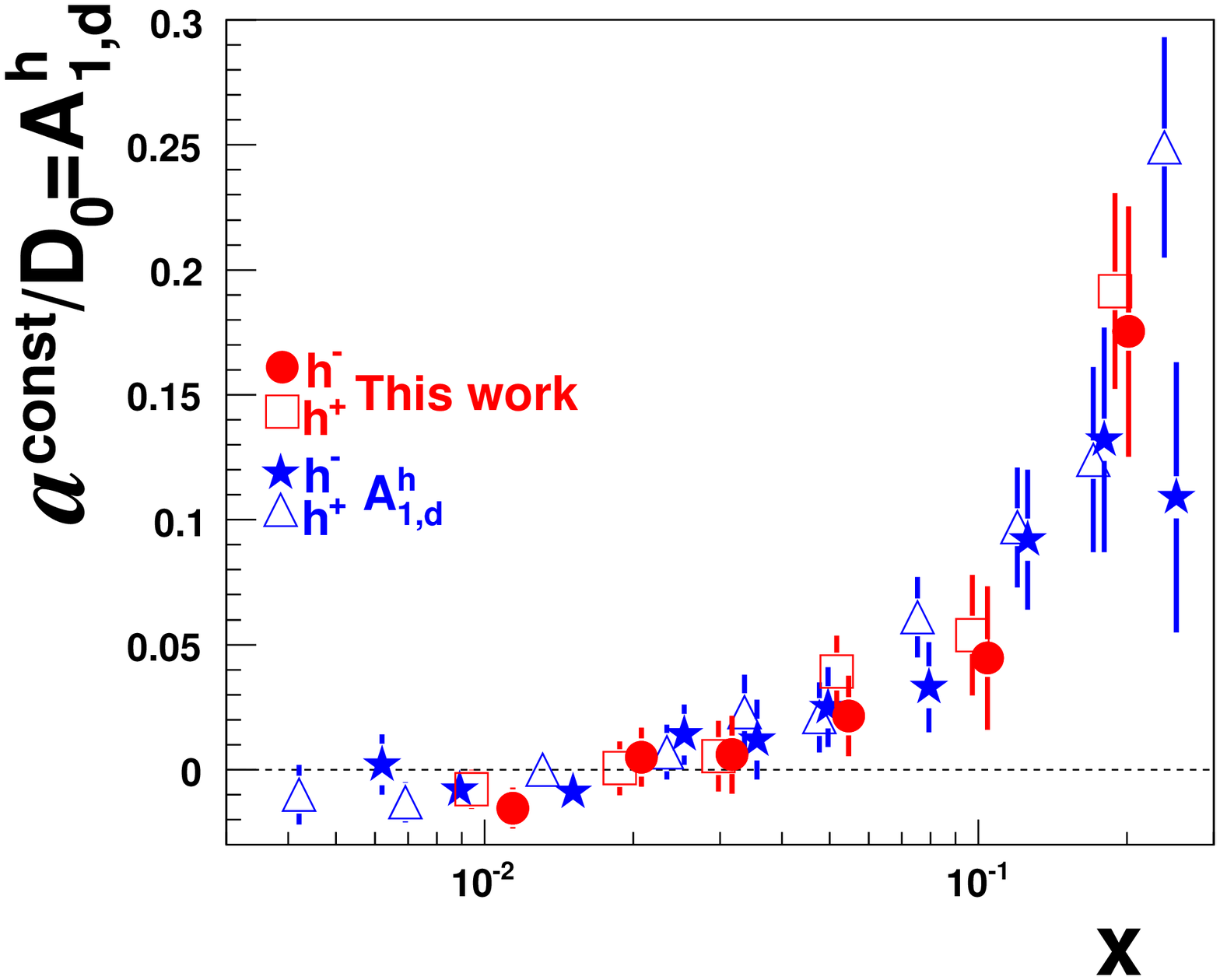}\hspace{-1mm}
\includegraphics[width=0.33\textwidth]
{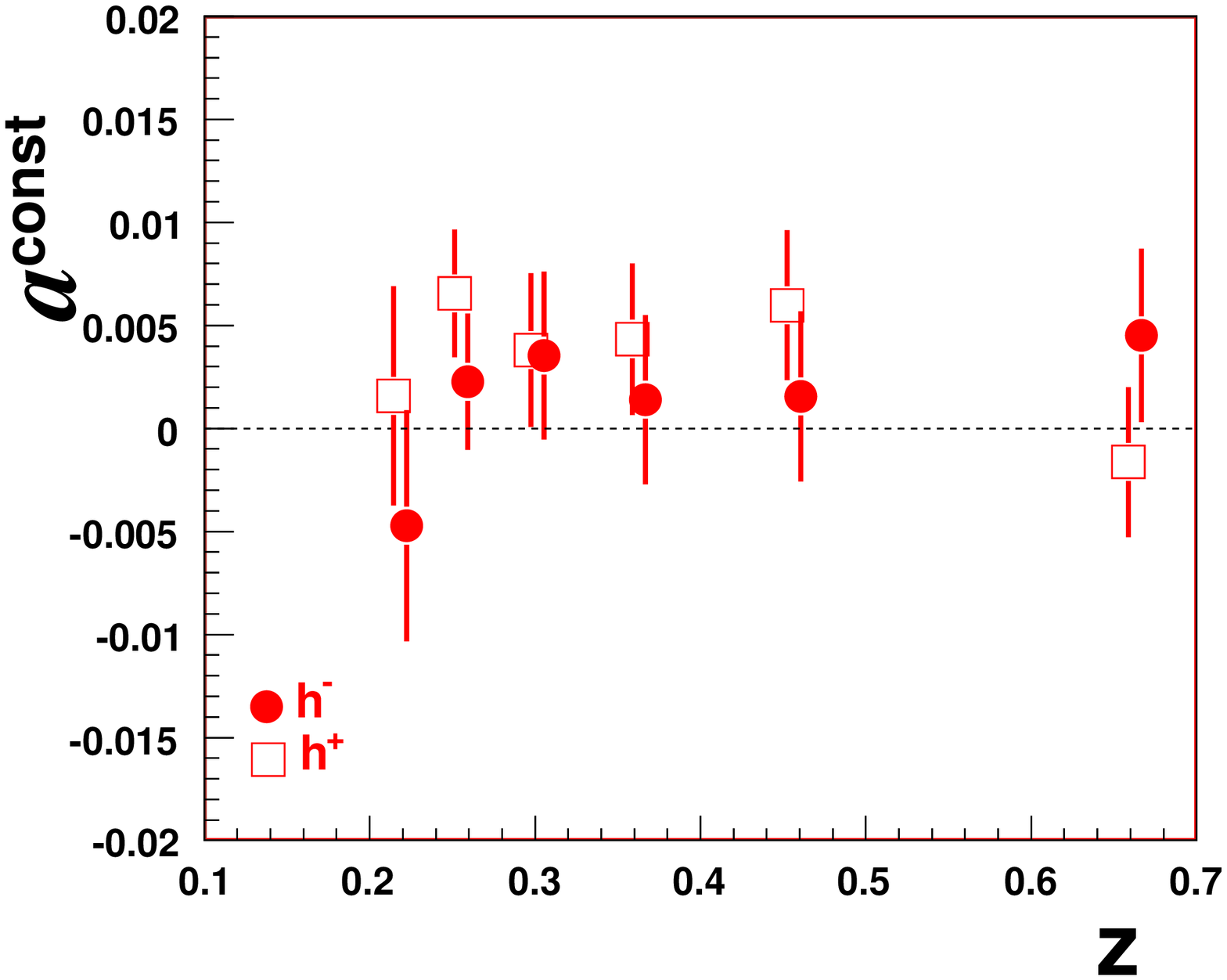}\hspace{-2mm}
\includegraphics[width=0.33\textwidth]
{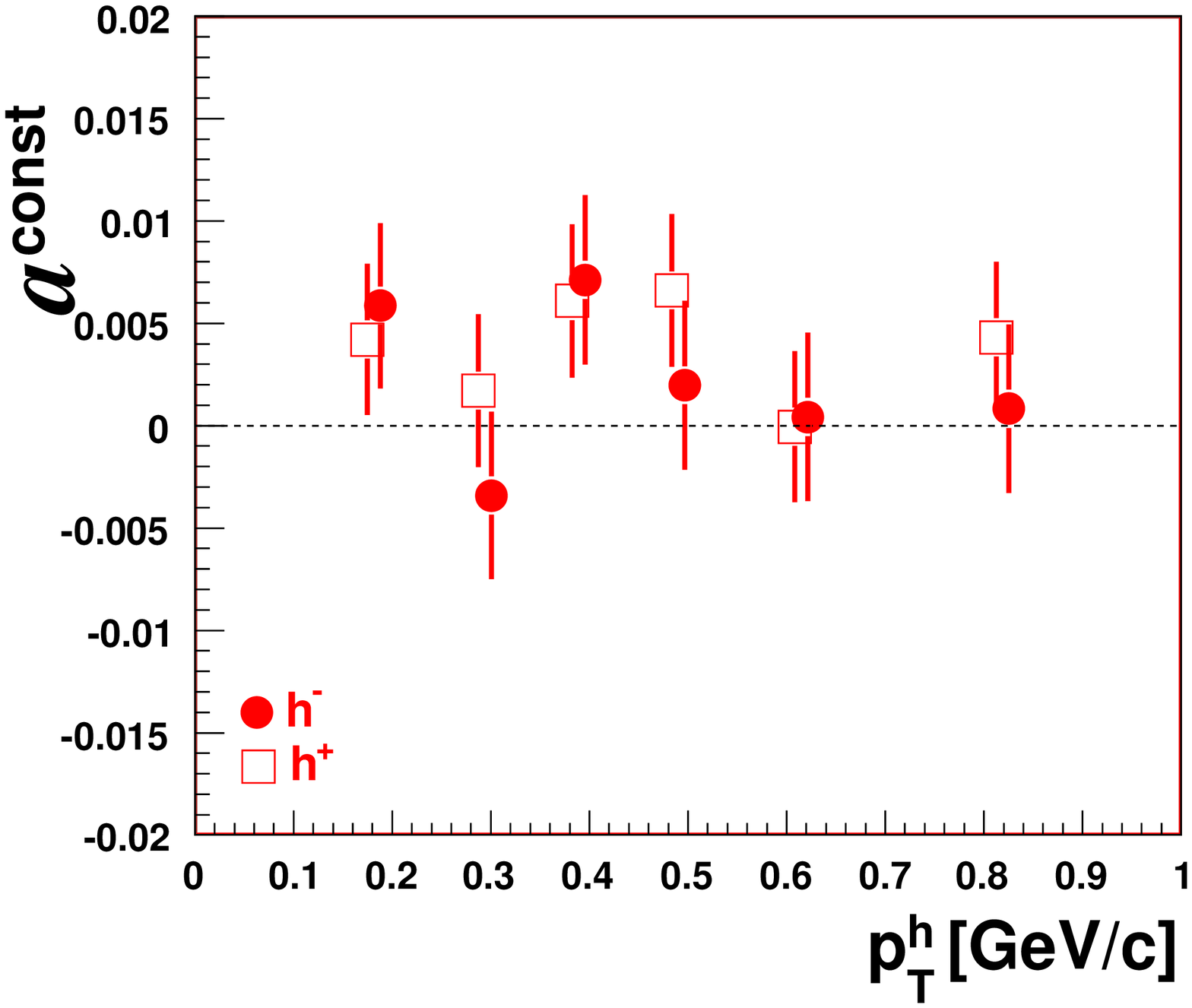} 
\caption
{\label{fig5}Dependence of the $a^{\rm const}$ parameter 
on the kinematic variables. The values of $A_{(1,d)}^h(x)$ from 
Ref.~\cite{Alekseev:2007vi} are also shown.}
\end{figure}
\begin{figure}[p]
\centering\vskip3mm  
\hspace{-3mm}
\includegraphics[width=0.32\textwidth]
{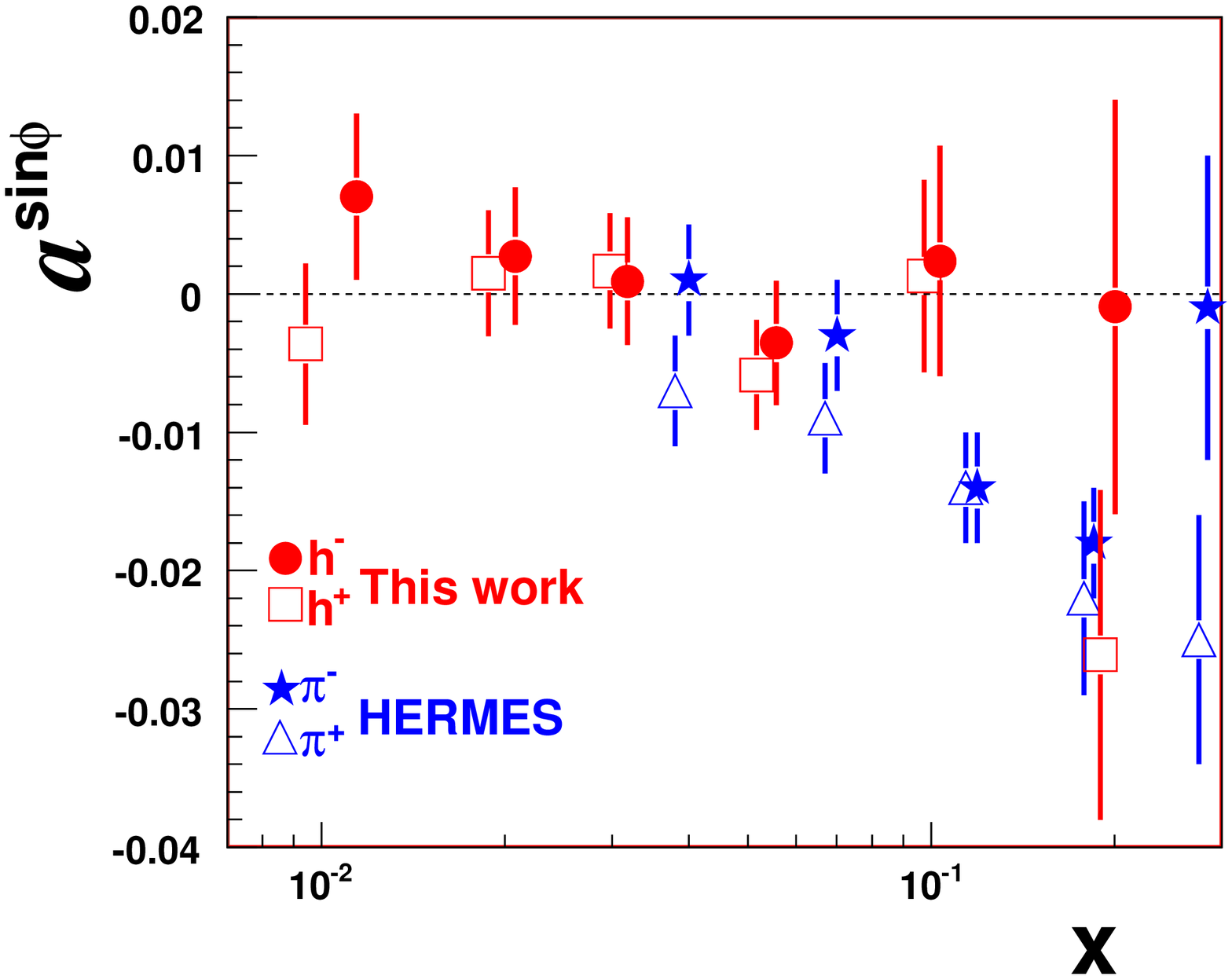}\hspace{-1mm}
\includegraphics[width=0.32\textwidth]
{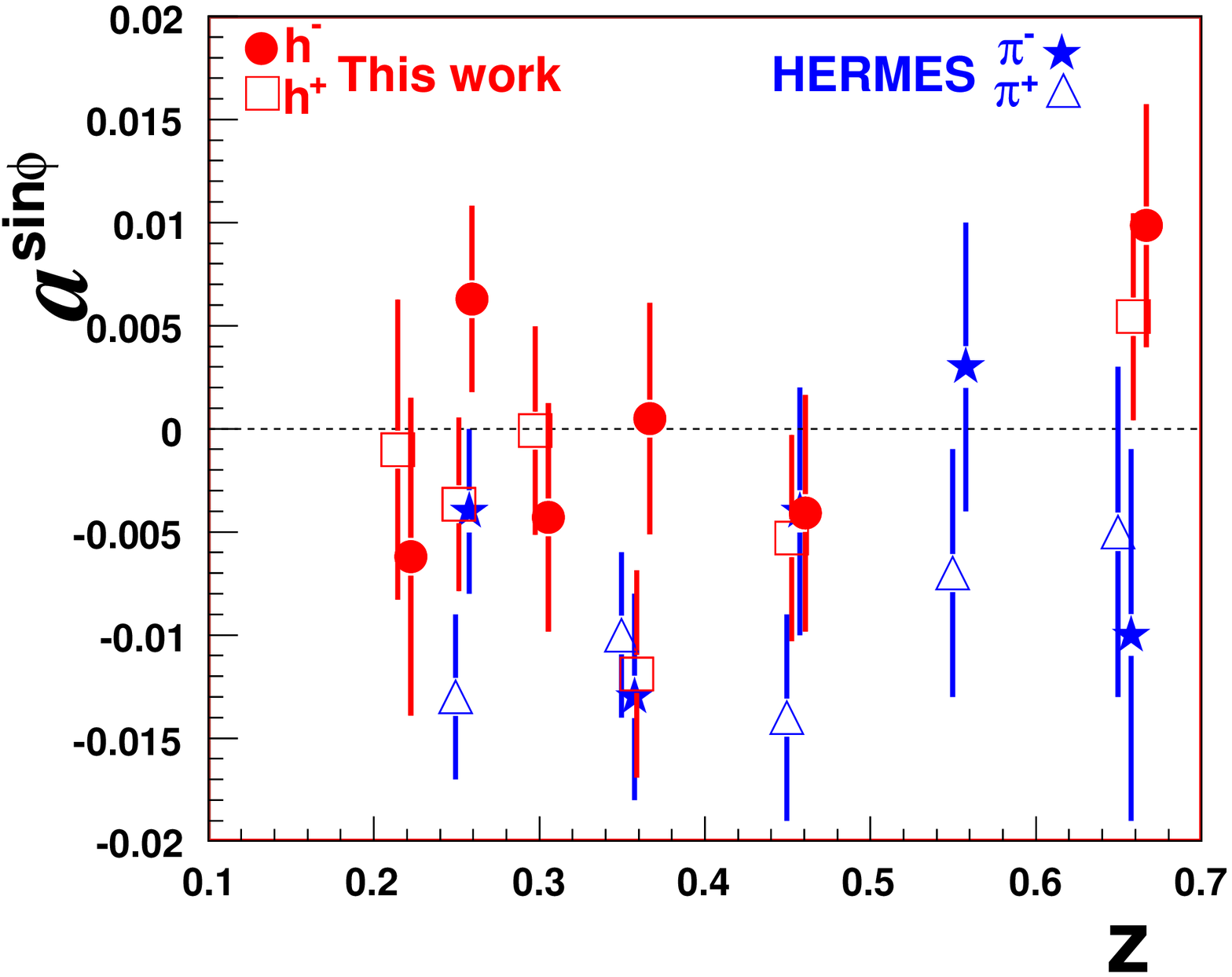}\hspace{-2mm}
\includegraphics[width=0.32\textwidth]
{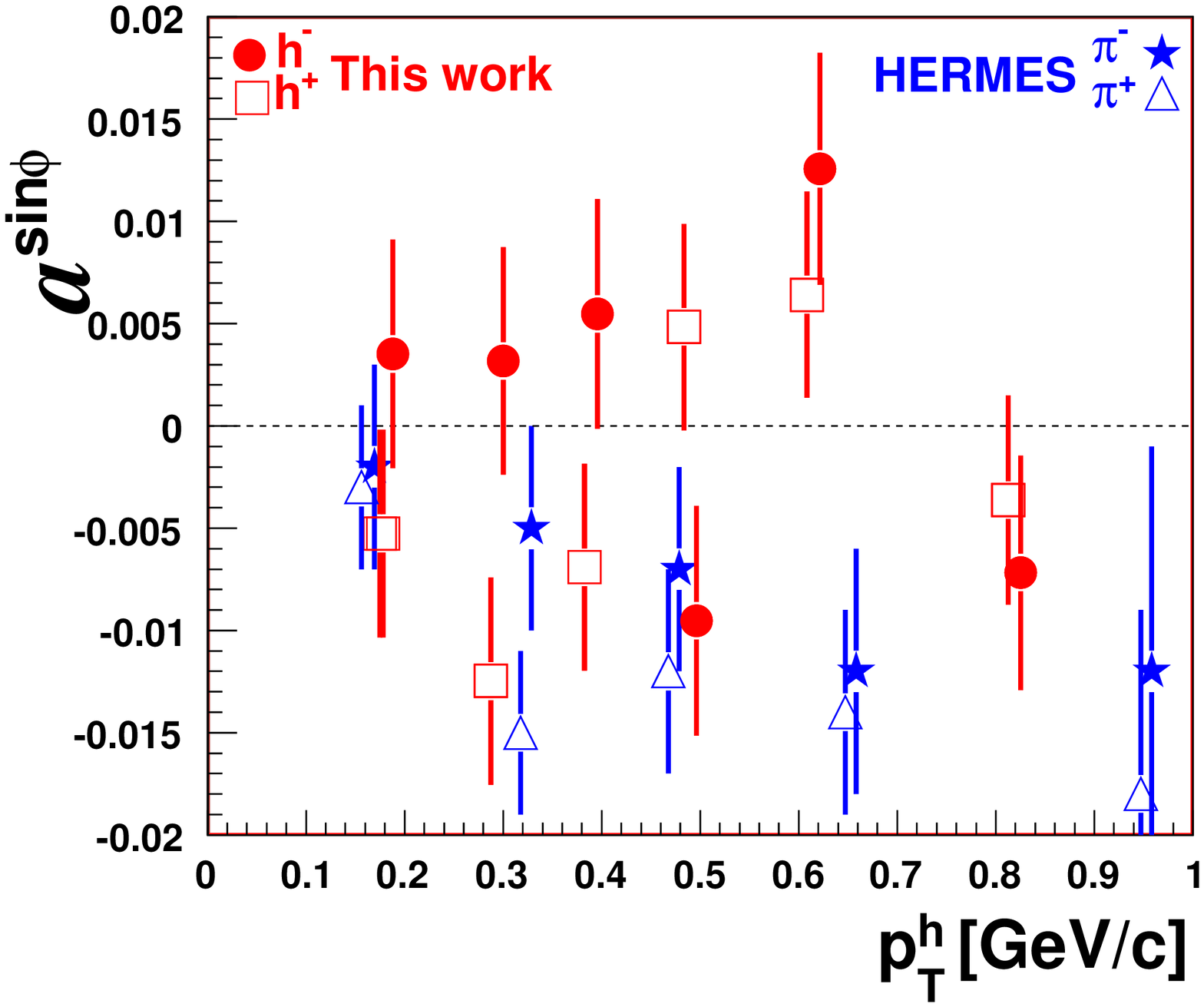} 
\caption
{\label{fig6}Dependence of the modulation amplitude 
$a^{\sin\phi}$ on the kinematic variables and similar data of 
HERMES \cite{Airapetian:2002mf} for identified leading pions.}
\end{figure}
\begin{figure}[p]
\centering 
\hspace{-3mm}
\includegraphics[width=0.34\textwidth]
{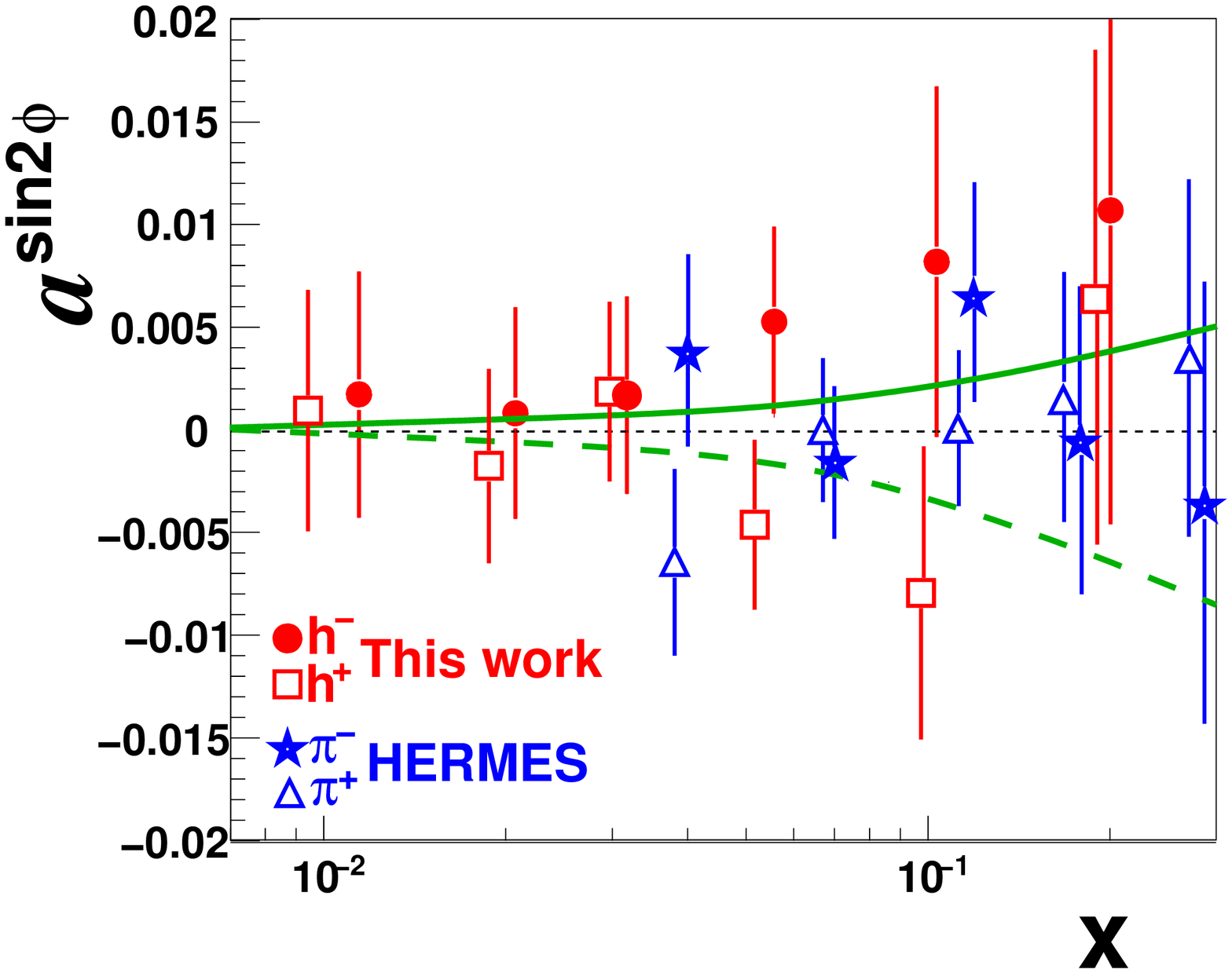}\hspace{-1mm}
\includegraphics[width=0.32\textwidth]
{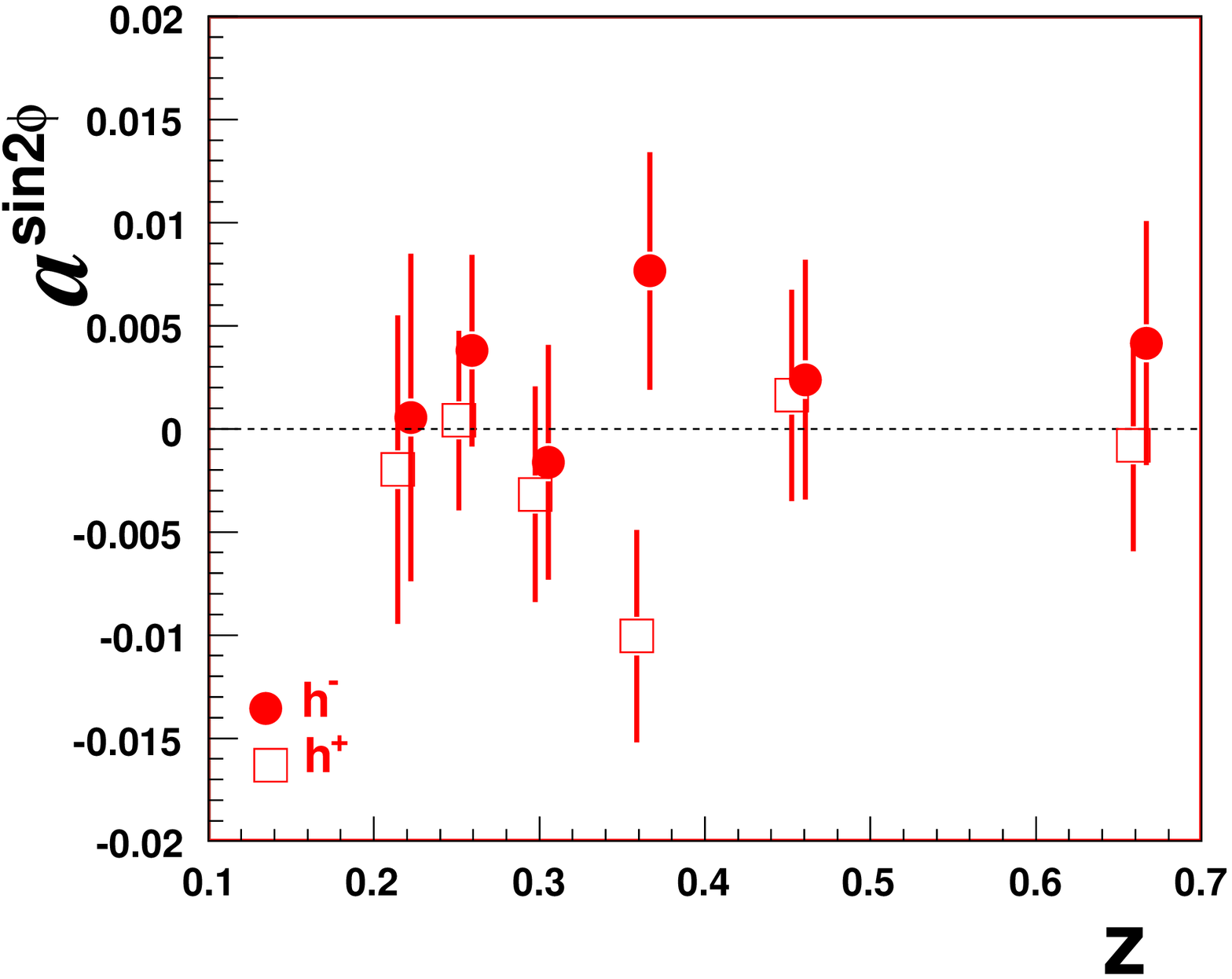}\hspace{-2mm}
\includegraphics[width=0.32\textwidth]
{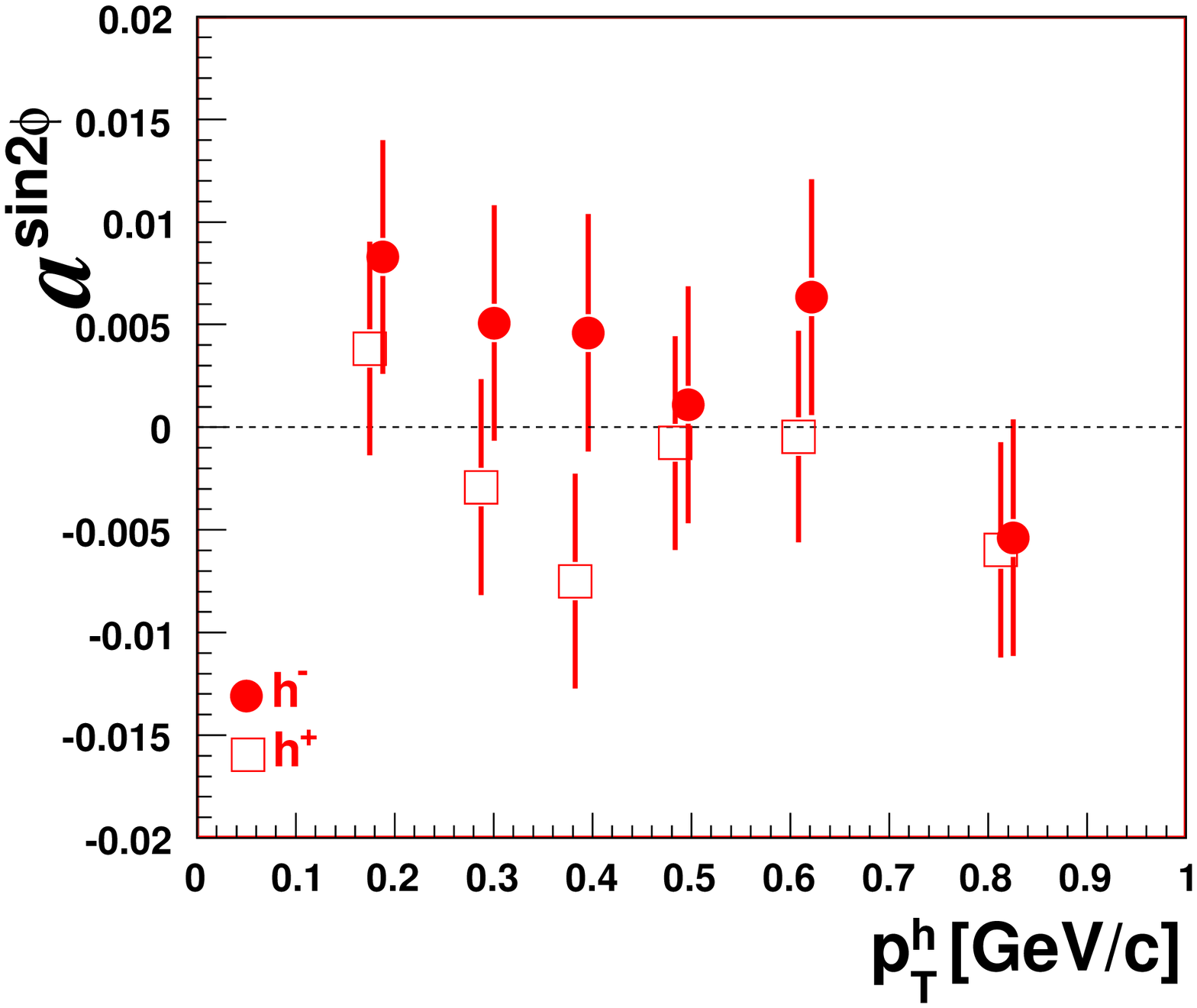}
\caption
{\label{fig7}Dependence of the modulation amplitude 
$a^{\sin2\phi}$ on the kinematic variables. The $x$ dependence of 
this amplitude is compared to the data and calculations by 
H.~Avakian {\em et al.} \cite{Avakian:2007mv} for HERMES 
kinematics$^{3)}$ for positive (solid line) and negative (dashed 
line) hadrons.}
\centering 
\hspace{-3mm}
\includegraphics[width=0.32\textwidth]
{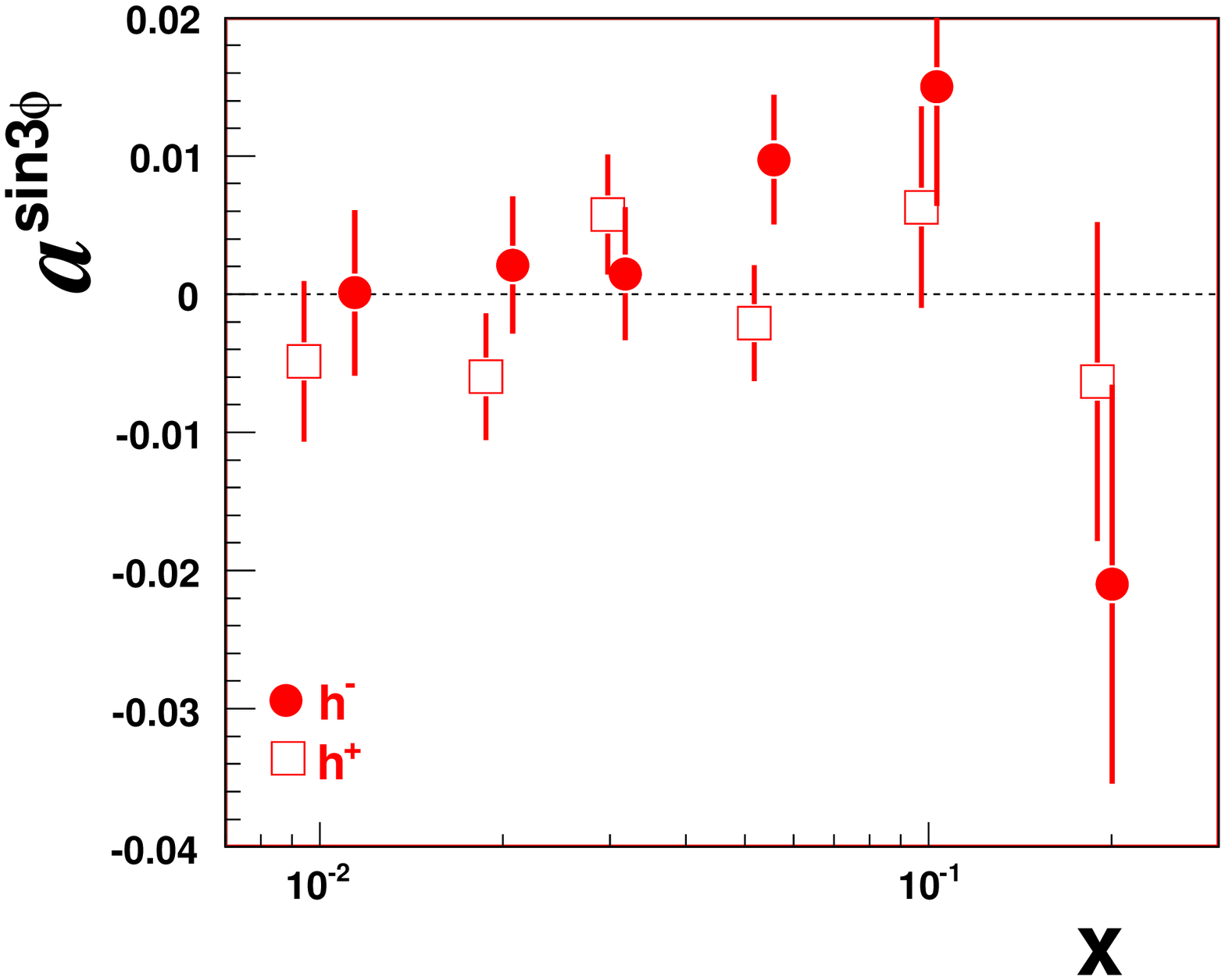}\hspace{-1mm}
\includegraphics[width=0.32\textwidth]
{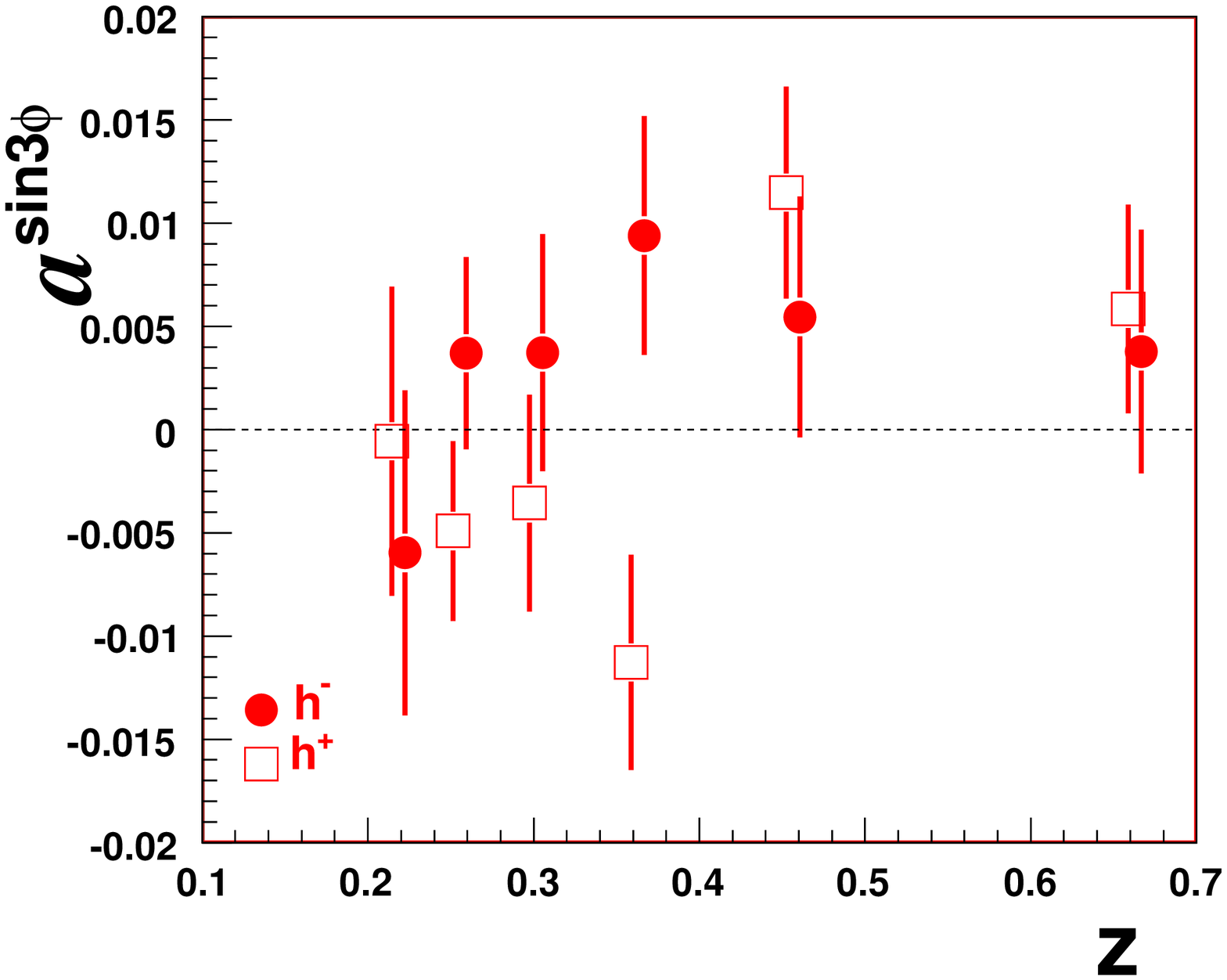}\hspace{-2mm}
\includegraphics[width=0.32\textwidth]
{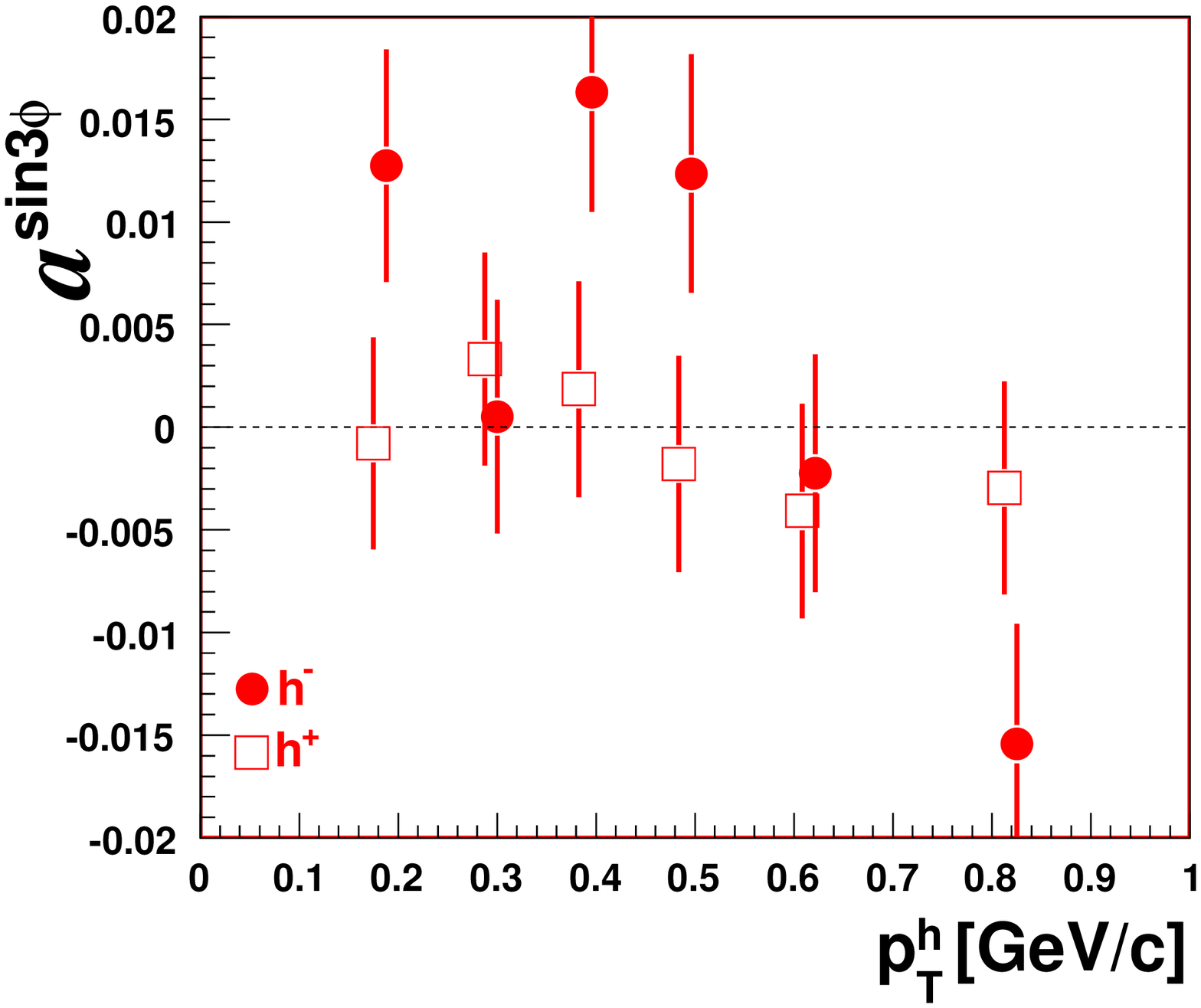}
\caption
{\label{fig8}Dependence of the modulation amplitude 
$a^{\sin3\phi }$ on the kinematic variables.}
\end{figure}
\begin{figure}[t]
\centering \hspace{-3mm}\vskip2mm
\includegraphics[width=0.32\textwidth]
{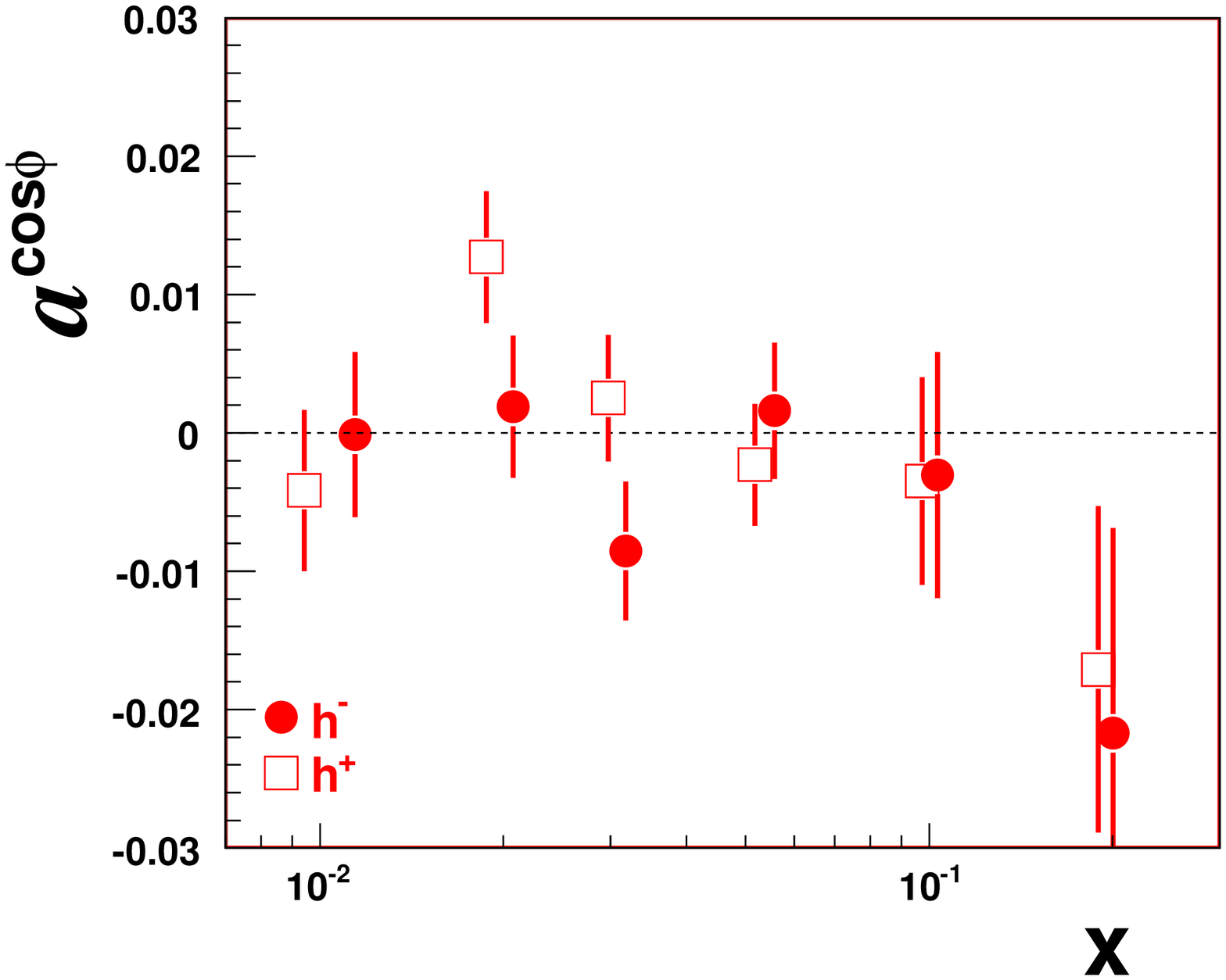}\hspace{-1mm}
\includegraphics[width=0.32\textwidth]
{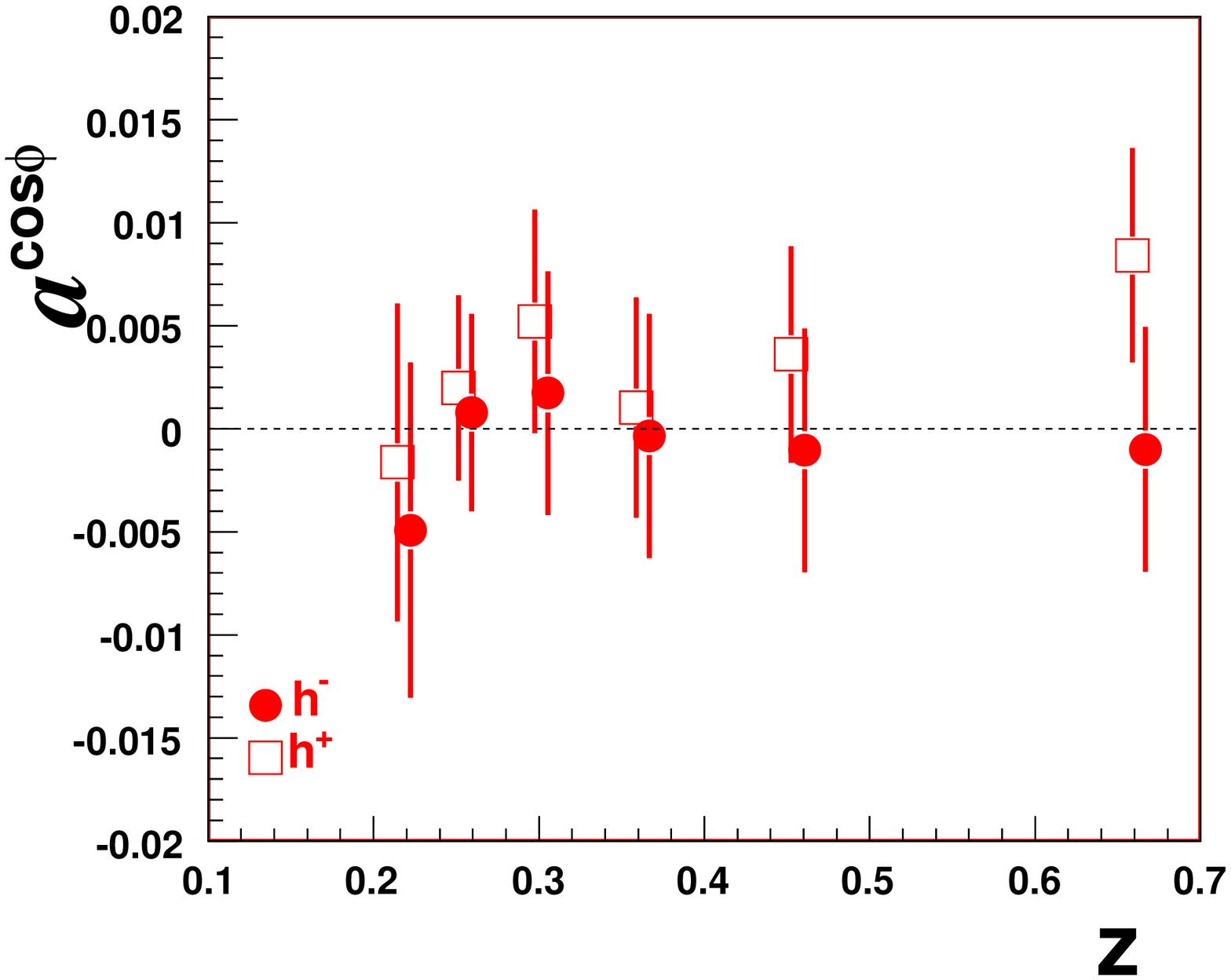}\hspace{-2mm}
\includegraphics[width=0.32\textwidth]
{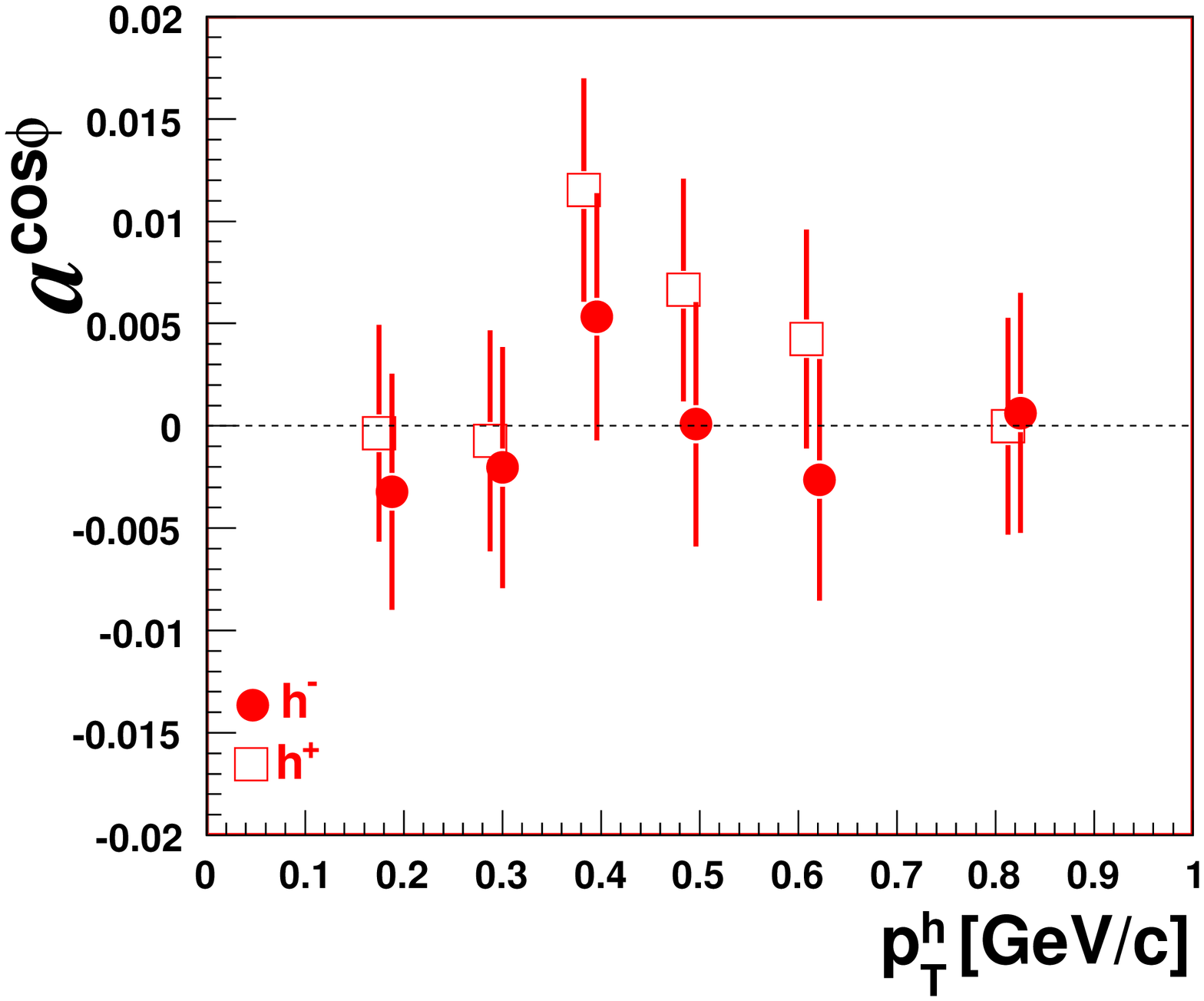}
\caption
{\label{fig9}Dependence of the modulation amplitude 
$a^{\cos\phi }$ on the kinematic variables.}
\end{figure}

The dependence of the amplitudes of the $\phi$ modulation on the 
kinematic variables is shown in Figs.~\ref{fig5}--\ref{fig9}. The 
bin sizes are optimised to contain more than $10^6$ events in 
each bin. Some points are slightly shifted horizontally for 
clarity. Only statistical errors are shown. Systematic 
uncertainties are estimated to be much smaller then the 
statistical ones (see Section~6).

The asymmetry parameter $a^{\rm const}(x)$ divided by the average 
muon polarisation and the virtual-photon depolarisation factor 
$D_{0}\approx |P_\mu|\sqrt{1-\epsilon^2}$ in the corresponding  
$x$ bin is shown in Fig.~\ref{fig5}. By definition, the value 
$a^{\rm const}(x)/D_0$ is equal to the asymmetry $A_{1,d}^h(x)$ 
published earlier \cite{Alekseev:2007vi}. The agreement of these 
data with those of the present analysis demonstrates the internal 
consistency of the results obtained by different methods.

The $\sin\phi$ modulation amplitudes of the azimuthal asymmetry 
are shown in Fig.~\ref{fig6}. Such a modulation is expected as a 
combined effect from the twist-3 PDFs $h_L$ and $f_L^\perp$ 
entering $\id\sigma_{0L}$ as well as from the twist-2 
transversity PDF $h_1$ and Sivers PDF $f_{1T}^\perp$ entering 
$\id\sigma_{0T}$ (see Eqs.~(\ref{eq3})), all contributing to the 
azimuthal asymmetries with a factor $Mx/Q$. The individual PDF 
contributions can not be separated within a single experiment. 
The observed $x$ dependence of this amplitude is less pronounced 
in the COMPASS data than in the HERMES \cite{Airapetian:2002mf} 
data.\footnote{The sign of HERMES data was inverted in order to 
match our definition of spin asymmetry by Eq.~(\ref{eq2}).} The 
latter are obtained for leading pions, while our data include all 
hadrons and cover a much wider range in $x$, $Q^2$ and $W^2$. 
When restricting our kinematic region to that of HERMES for the 
amplitude $a^{\sin\phi}$ compatible results are obtained.

The amplitudes of the $\sin(2\phi)$ modulation shown in 
Fig.~\ref{fig7} are small, consistent with zero within the 
errors. They are due to the PDF $h_{1L}^\bot$ in $\id\sigma_{0L}$ 
(see Eqs.~(\ref{eq3})) which is approximately linked 
\cite{Avakian:2007mv} to the transversity PDF $h_1$ by a  
relation of the Wandzura--Wilczek type.

The data on the modulation amplitude $a^{\sin3\phi}$ shown in 
Fig.~\ref{fig8} are compatible with zero as our results on the 
amplitude of the $\sin(3\phi-\phi_S)$ modulation extracted from 
transversely polarised deuterons \cite{Kotzinian:2007uv} ($\phi_S 
= 0$, $\pi$ for longitudinal target polarisation). This 
modulation would be due to the pretzelosity PDF $h_{1T}^\bot$ in 
$\id\sigma _{0T}$ and is suppressed by the factor 
$\tan\theta_\gamma\sim xM/Q$.

The $\cos\phi$ modulation of the azimuthal asymmetries for a
longitudinally polarised target is studied here for the first 
time. The data presented in Fig.~\ref{fig9} are consistent with 
no variations of the modulation amplitudes vs.\ $x$, $z$ and 
$p_T^h$. This amplitude is proportional to the muon beam 
polarisation and would be due to a pure twist-3 PDF $g_L^\bot$ in 
$\id\sigma_{LL}$, an analogue to the Cahn effect 
\cite{Cahn:1978se} in unpolarised SIDIS \cite{Arneodo:1986cf}, 
and $g_{1T}$ in $\id\sigma_{LT}$, suppressed by 
$\tan\theta_\gamma\sim xM/Q$.
 
\section{Stability of the results and systematics} 
\begin{figure}[t]
\centering
\hspace{-3mm}
\includegraphics[width=0.33\textwidth]
{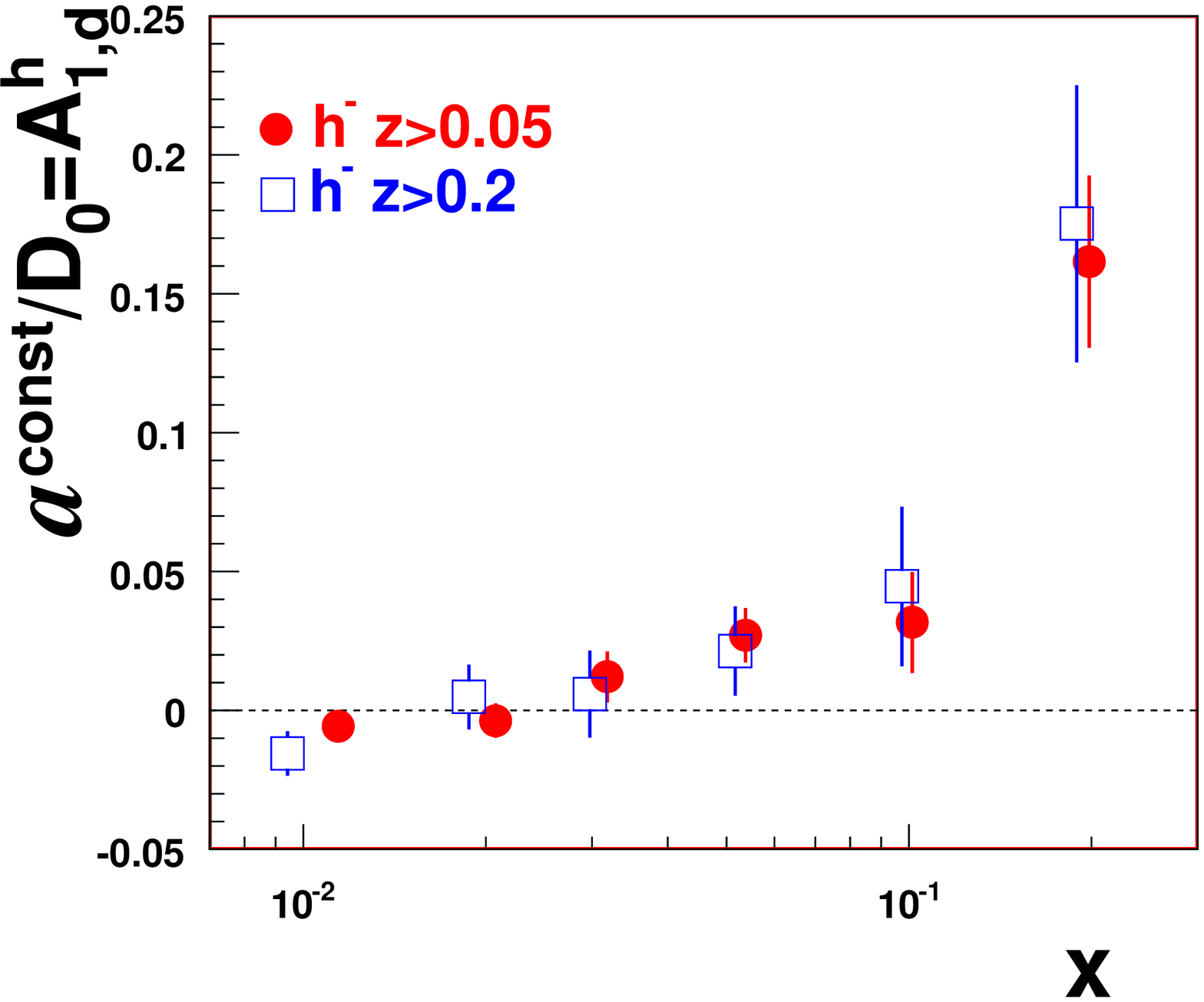}\hspace{-2mm}
\includegraphics[width=0.33\textwidth]
{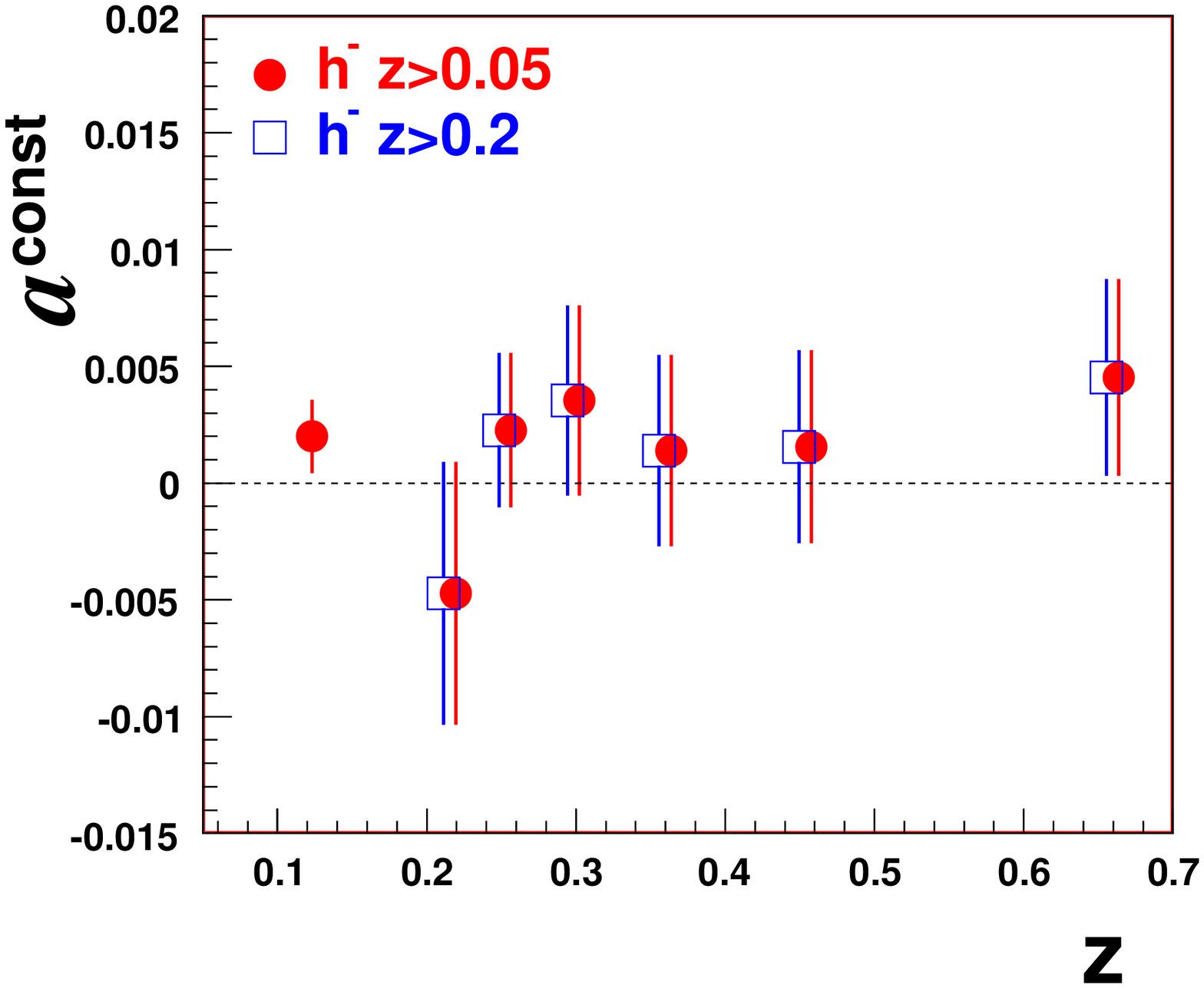}\hspace{-2mm}
\includegraphics[width=0.33\textwidth]
{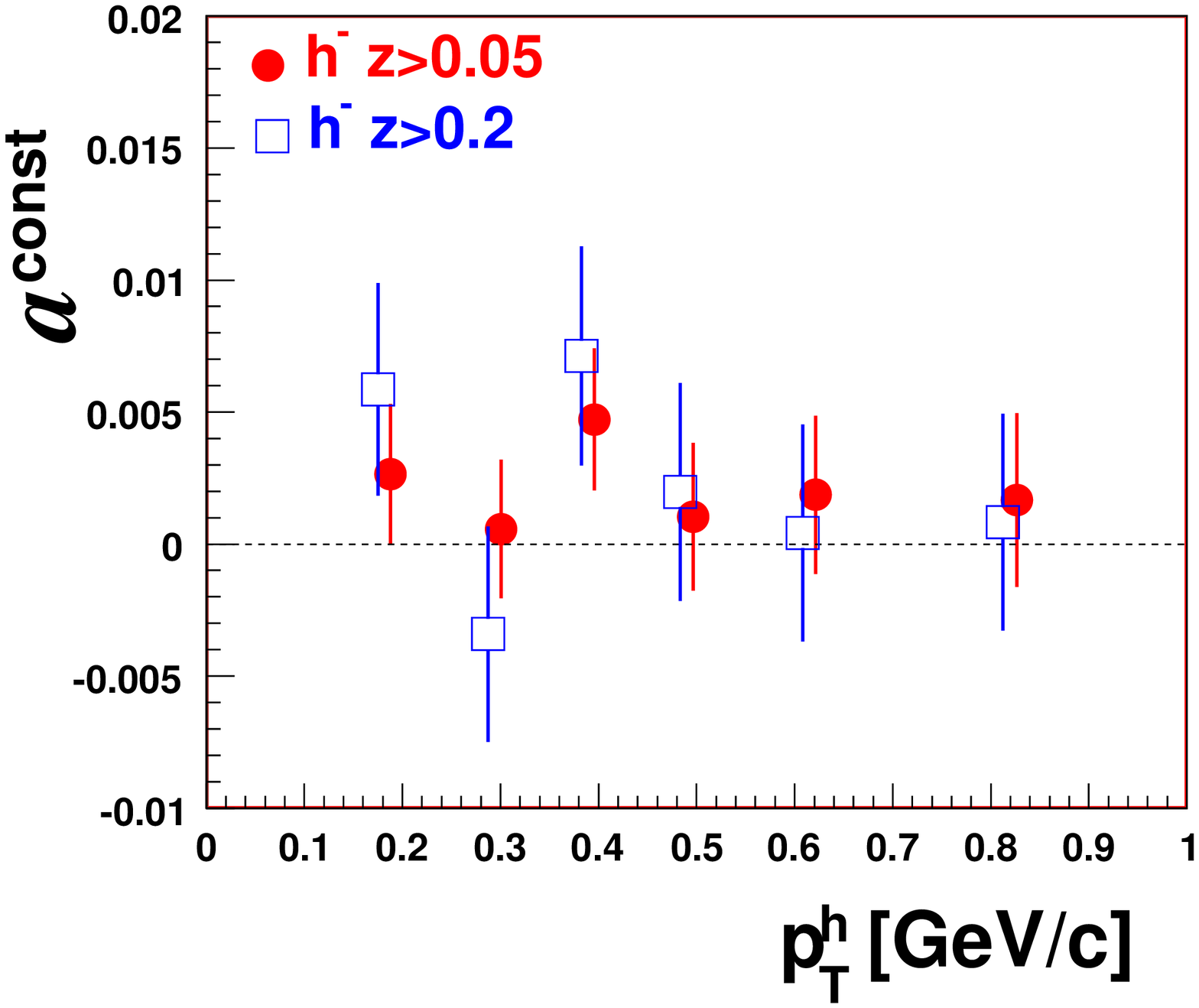}
\caption
{\label{fig5a}Comparison of results for the $h^-$ 
azimuthal asymmetry parameter $a^{\rm const}(x)/D_0$, $a^{\rm 
const}(z)$ and $a^{\rm const}(p_T^h)$ for different regions: $z=$ 
0.05--0.9 and $z=$ 0.2--0.9.}
\end{figure}

%
In the above analysis the $z$ cut ($z>0.2$) has been applied to 
reject hadrons originating from the target fragmentation region. 
To check if a lower cut can affect the results presented in Table 
1 and in Figs.~\ref{fig5}--\ref{fig9}, we have repeated the 
analysis with the cut $z>0.05$ and obtained consistent results. 
The results for the two different $z$ cuts are compared in 
Fig.~\ref{fig5a} for the asymmetry parameters $a^{\rm 
const}(x)/D_0$, $a^{\rm const}(z)$, and $a^{\rm const}(p_T^h)$.
The $x$ dependence of this parameter does not 
depend on the applied $z$ cut, i.e.\ there is no 
influence of the target fragmentation region on the $x$ dependence for the
selected sample of events down to $z=0.05$.
The observed $p_T^h$ dependence looks smoother with higher statistics.

\begin{figure}[t]
\centering
\includegraphics[width=0.60\textwidth]
{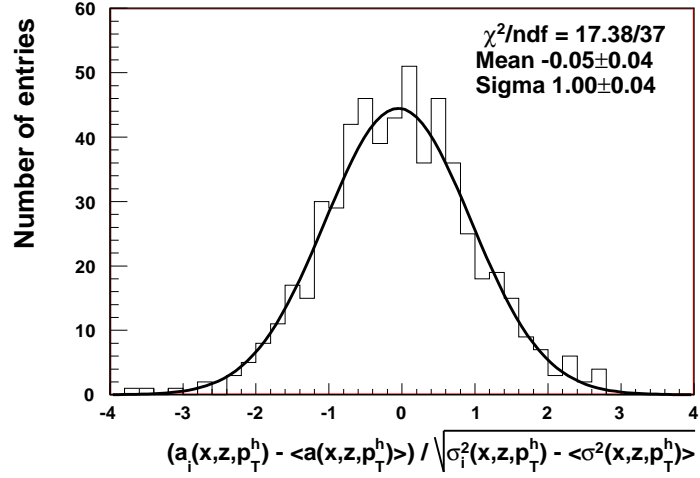}
\caption
{\label{fig11}Distribution of pulls for measurements 
of $a_i$.}
\end{figure}

%
The compatibility of the final  results on the azimuthal 
asymmetries obtained with the data taken in 2002--2004 has been
checked by building distributions of ``pulls"
\begin{equation}
\label{eq10}
\frac{a_i-\langle a\rangle}{\sqrt{\sigma_{a_i}^2-
\sigma_{\langle a\rangle}^2}} 
\end{equation} 
where $a_i$ is the asymmetry in a given year for a given hadron 
charge and bin of the kinematic variables $x$, $z$, and $p_T^h$. 
The value $\langle a\rangle$ is the corresponding weighted mean 
over three years and $\sigma$ is the statistical error. The 
overall distribution of pulls for all $a_i$ measurements (540 
entries, i.e.\ 5 asymmetries for positive and negative hadrons, 3 
variables with 6 bins and 3 years of data taking) is shown in 
Fig.~\ref{fig11}.  
As expected, the distribution of pulls follows a 
standard distribution with mean close to zero and sigma 
equal to unity indicating that the fluctuations in the data are only 
statistical.

The double ratio Eq.~(\ref{eq4}) has been used to extract the 
asymmetries because of the cancellation of acceptance and flux. 
To check this cancellation, two different double ratios have been 
constructed using the same number of events, namely
\begin{equation}
\label{eq11} F_+(\phi)=\frac{N_{++}^U N_{-+}^D}{N_{++}^D N_{-+}^U}
\end{equation}     
and 
\begin{equation} 
\label{eq12} F_-(\phi)=\frac{N_{+-}^U 
N_{--}^D}{N_{+-}^D N_{--}^U}
\end{equation}
for the positive and negative solenoid field orientations, 
respectively. Substituting into Eqs.~(\ref{eq11}, \ref{eq12}) the 
number of events by the corresponding expressions given in 
Eq.~(\ref{eq5}), one could see that these ratios depend neither 
on the acceptance nor on the DIS cross-section. The above 
equations should reflect the relative integrated muon fluxes 
squared for the measurements with different target polarisation 
but the same solenoid field orientation and they are a good check 
of acceptance cancellation in Eq.~(\ref{eq4}).
\begin{figure}[th]
\centering 
\vskip-3mm
\includegraphics[width=0.7\textwidth]
{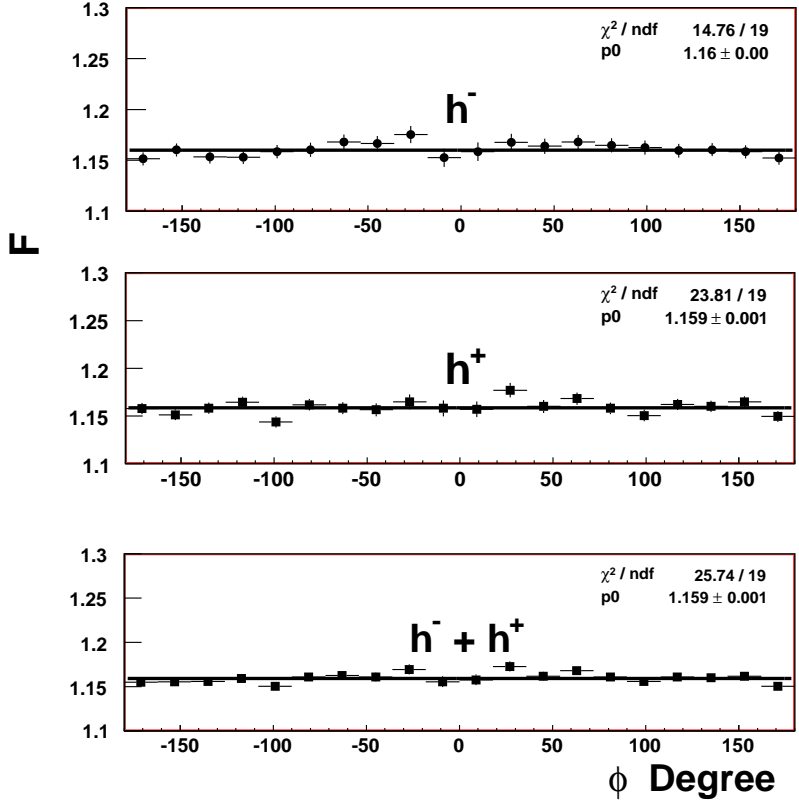}
\caption
{\label{fig12} 
Double ratios of event numbers for the 
weighted sum $F=F_+(\phi)\oplus F_-(\phi)$ demonstrating the 
$\phi$ stability of the data taking rates for negative (top), 
positive (middle) and all hadrons (bottom). The parameters $p_0$ 
represent the results of fits by a constant. }
\end{figure}

Indeed, the ratios Eq.~(\ref{eq11}, \ref{eq12}) for the selected 
events integrated over all kinematic variables are found to be 
independent of $\phi$ and the field orientations and are the same 
for $h^+$ and $h^-$ (i.e., $\phi$-dependent acceptances 
$C_f^t(\phi)$ are really cancelled). This is illustrated in 
Fig.~\ref{fig12} where the weighted sums of $F=F_+(\phi)\oplus 
F_-(\phi)$ for negative and positive hadrons are shown together 
with results of their fit by a constant.


The compatibility and stability tests indicated no systematic 
effects. It is therefore assumed that the systematic 
uncertainties of the measured asymmetries due to variations of 
the acceptance are smaller than the statistical errors.

The uncertainties due to the target and beam polarisation are 
estimated to be of the order of 5\% each. The uncertainty of the 
dilution factor, which takes into account the target material 
composition, is of the order of 2{\%} \cite{Ageev:2005gh}. Each 
of them introduces the corresponding multiplicative uncertainty 
in the asymmetry measurement. When combined in quadratures, these 
errors give a global systematic multiplicative uncertainty of 
less than 7\%. The errors due to the additive radiative 
correction are negligible in the used kinematical region.

\section{Conclusions}
Azimuthal asymmetries $a(\phi)$ were studied in the 
production of positive and 
negative hadrons by 160~GeV muons scattering off the 
longitudinally polarised deuterons. Integrated over 
the variables $x$, $z$, and $p_T^h$, all $\phi$-modulation 
amplitudes of $a(\phi)$ are consistent with zero, while the 
$\phi$-independent parts of the $a(\phi)$ distributions differ 
from zero and are equal for positive and negative hadrons within 
the statistical errors. In the study of $a(\phi)$ as a function 
of $x$, $z$, and $p_T^h$ the following results are obtained. The 
$\phi$-independent terms $a^{\rm const}(x)$ of $a(\phi)$ are in 
agreement with our earlier results \cite{Alekseev:2007vi} and 
other published data on $A^h_{1,d}$, calculated by other methods 
and using different cuts. The amplitudes 
$a^{\sin\phi}$ are small and in general compatible with the 
HERMES data \cite{Airapetian:2002mf}. The amplitudes 
$a^{\sin2\phi}(x,z,p_T^h)$, $a^{\sin3\phi}(x,z,p_T^h)$ and 
$a^{\cos\phi}(x,z,p_T^h)$ are consistent with zero within 
statistical errors. 

These data will be useful to constrain models of the nucleon 
structure. The present parton model description of the SIDIS 
cross-sections involves a considerable number of PDFs depending 
on the longitudinal and transverse components of the nucleon 
spin. We believe that our data will help to assess which PDFs are 
important in the description of the nucleon structure.

\subsection*{Acknowledgements} 
We gratefully acknowledge the support of the CERN management and 
staff and the skill and effort of the technicians of our 
collaborating institutes. Special thanks go to V.~Anosov and 
V.~Pesaro for their technical support during the installation and 
running of this experiment.

\end{document}